\documentclass{article}
\usepackage{amssymb,latexsym}
\usepackage{amsmath}

\begin{document}
\title{Time as an operator/observable in\\nonrelativistic quantum mechanics}        
\author{G. E. Hahne\thanks{email:  hahne@nas.nasa.gov}\\
        NASA, Ames Research Center\\
        Moffett Field, California, 94035 USA}
\maketitle

\centerline{PACS Numbers:  03.65.Ca, 03.65.Xp, 02.30.Jr}
\vskip 5pt
\begin{abstract}
     The nonrelativistic Schr\"odinger
equation for motion of a structureless particle in four-dimensional
space-time entails a well-known expression for the 
conserved four-vector field of local probability density and current
that are associated with a quantum state solution to the equation.
Under the physical assumption that each spatial, as well as the
temporal, component of this current is observable,
the position in time becomes an operator and an observable in that the weighted
average value of the time of the particle's crossing of a complete
hyperplane can be simply defined:  the theory predicts,
and experiment is presumed to be able to observe, the integral
over the hyperplane of the normal component of probability current, weighted by
the time coordinate.  In conventional formulations the
hyperplane is always spacelike, i.e., is a time=constant
hyperplane in Galilean relativity, and the result is then trivial.
A nontrivial result is obtained if the plane is not of
this type.  When the space-time coordinates are $(t,x,y,z)$,
the paper analyzes in detail the case that the hyperplane
is of the type $z$=constant.  Particles can cross such a hyperplane
in either direction, so it proves convenient to introduce
an indefinite metric, and correspondingly a sesquilinear inner product
with non-Hilbert space
structure, for the space of quantum states on such a surface.
Since the metric is indefinite, an uncertainty principle
involving the dispersion of the crossing time and 
the dispersion of its conjugate momentum does not appear
to be derivable from the theory.  A detailed formalism for computing
average crossing times on a $z$=constant hyperplane, 
and average dwell times and delay times for a zone of interaction
contained between a pair of $z$=constant hyperplanes, is presented.

\end{abstract}


\section{Introduction} \label{S:sec1}

     Within a few years after the discovery of quantum mechanics  
a consensus formed 
(von Neumann \cite{R:Neumann2}, p.\ 188 and \cite{R:Neumann1}, p. 354,
Pauli \cite{R:Pauli2}, p.\ 140, footnote, 
and \cite{R:Pauli1}, p.\ 63, footnote) 
to the effect that, in contrast to spatial positions, and therefore
in conflict with special relativity, 
the temporal position $t$ is necessarily a c-number, or parameter,
with no generic operator status being mathematically feasible.
In the decades intervening since the publication of the original 
versions of the two cited treatises in 1932 and 1933, 
respectively, the prohibition on specifying the time
as a dynamical variable has been widely upheld as part of
the standard doctrine of quantum mechanics---see, e.\ g.,
Peres \cite{R:Peres1} Chap.\ 12-7, Omn\`es \cite{R:Omnes1}  p. 57,
and Sakurai \cite{R:Sakurai1} p.\ 68.
In recent decades interest in this subject has intensified,
due in part to applications of tunneling phemomena in semiconductors
(\cite{R:Capasso1}, \cite{R:Ferry1} Ch.\ 3.2.3, \cite{R:Jonson1},
\cite{R:Jauho1}, \cite{R:Jauho2}, \cite{R:Mizuta1}),
and a substantial set of results has been published that 
introduce formalisms that argue for, or against, 
various quantum-mechanical definitions of time, 
including arrival times, tunneling times, dwell (alias sojourn)
times, and delay times.
For definitions, reviews, and citations, see Refs.\ 
\cite{R:Hauge1}, \cite{R:Olkhovsky1}, \cite{R:Landauer1},
\cite{R:Chiao1}, \cite{R:Mugnai1}, \cite{R:Damborenea1},
and a collection of articles in \cite{R:Muga0}.

    This paper claims to advance a general operator for time
as a quantum-mechanical observable in the context of the
nonrelativistic Schr\"odinger equation.
A natural preliminary question is, how does the construction of 
such an
observable time square with the putative result that no such operator
exists in general?  I believe that the nonexistence proof
rests on an improper substitution of dynamics for kinematics, 
and that there is a straightforward interplay between classical and quantum
mechanics on this question.  We infer from Pauli \cite{R:Pauli2}
in a footnote on p.\ 140, that
the search for a classical observable for time reduces to
the following problem: 
Let $\{q^1,p_1,\ldots,q^N,p_N\}$ be the positions and
momenta in a generic classical Hamiltonian dynamics, with Hamiltonian
$H(t,q^1,p_1,\ldots,q^N,p_N)$.  We define the ordinary Poisson bracket
$\{f_1,f_2\}_{\text{pb}}$ of two functions $f_{1,2}(t,q^1,p_1,\ldots,q^N,p_N)$ 
to be
\begin{equation}\label{E1:eq1}
\{f_1,f_2\}_{\text{pb}}\ =\ \sum_{j=1}^N
\biggl(\frac{\partial f_1}{\partial q^j}\frac{\partial f_2}{\partial p_j}- 
\frac{\partial f_1}{\partial p_j}\frac{\partial f_2}{\partial q^j}\biggr),
\end{equation}
Then we want to find a function $T(t,q^1,p_1,\ldots,q^N,p_N)$
such that
\begin{equation}\label{E1:eq1A}
\{T,H\}_{\text{pb}}\ =\ 1.
\end{equation}
Such a function $T$ would be the classical limit of a quantum-mechanical
Hermitean operator $T_{\text{qm}}$ that is invoked by Pauli, but demonstrated
by him not to exist in general,
such that the commutator satisfies
\begin{equation}\label{E1:eq1B}
[T_{\text{qm}},H_{\text{qm}}]\ =\ i\hbar 1_{\text{qm}}.
\end{equation}
Pauli's theorem has prevailed over the intervening decades,
and attempts to define an observable time in connection with solutions
to the Schr\"odinger equation have had recourse to many
alternate approaches---see the reviews cited above.

   It is easy to see in a related context that Pauli's
result, although correct in its limited context, 
addresses the wrong question.  
Let us define the augmented Poisson bracket
$\{F_1,F_2\}_{\text{apb}}$ of two functions
$F_{1,2}(t,p_t,q^1,p_1,\ldots,q^N,p_N)$ to be
\begin{equation}\label{E1:eq2}
\{F_1,F_2\}_{\text{apb}}\ =\ 
\frac{\partial F_1}{\partial t}\frac{\partial F_2}{\partial p_t}-
\frac{\partial F_1}{\partial p_t}\frac{\partial F_2}{\partial t}
+\{F_1,F_2\}_{\text{pb}}.   
\end{equation}
The function $\mathcal{D}_{\text{cl}}$, defined as
\begin{equation}\label{E1:eq3}
\mathcal{D}_{\text{cl}}\ =\ p_t+H,
\end{equation}
generates the time dependence of quantities as $F$, in that
\begin{equation}\label{E1:eq4}
\frac{dF}{dt}\ =\ \{F,\mathcal{D}_{\text{cl}}\}_{\text{apb}},
\end{equation}
subject to the dynamical constraint that on classical paths we require
\begin{equation}\label{E1:eq5}
\mathcal{D}_{\text{cl}}\ =\ 0
\end{equation}
(see \cite{R:Lanczos1}, Ch.\ VI.10).
We now infer that if
$F=t$, then $dF/dt=\{t,\mathcal{D}_{\text{cl}}\}_{\text{apb}}=1$,
whatever be the Hamiltonian $H$.  In quantum mechanics, with 
suitable attention to operator ordering, we substitute
\begin{equation}\label{E1:eq6}
\mathcal{D}_{\text{qm}}\ =\ \frac{\hbar}{i}\frac{\partial}{\partial t}
+H(t,q^1,\frac{\hbar}{i}\frac{\partial}{\partial q^1},\ldots),
\end{equation}
and obtain the time-dependent Schr\"odinger equation for
$\psi(t,q^1,\ldots,q^N)$ as a dynamical constraint on $\psi$, that is,
\begin{equation}\label{E1:eq7}
\mathcal{D}_{\text{qm}}\psi\ =\ 0.
\end{equation}
If $\psi$ is a solution to the Schr\"odinger equation we can substitute
$H\psi$ for $i\hbar\partial\psi/\partial t$, but not otherwise.
In particular, if $\psi$ is a nontrivial solution to \eqref{E1:eq7},
$t\psi$ is not a solution, so that
\begin{equation}\label{E1:eq8}
i\hbar\frac{\partial}{\partial t}t\psi\ \neq\ Ht\psi.
\end{equation}
Therefore, trying to satisfy the operator commutation rule \eqref{E1:eq1B}
is not relevant to the problem of finding
an operator for the time, even when the operand is a solution
to the Schr\"odinger equation.  However, the general
definition for the time-derivative of quantum-mechanical operators
\begin{equation}\label{E1:eq9}
\frac{dF_{\text{qm}}}{dt}\ =\ (i\hbar)^{-1}
[F_{\text{qm}},\mathcal{D}_{\text{qm}}]
\end{equation}
does make sense no matter what the state function,
dynamically constrained or not, operated on by the rhs:  
in fact, if $F_{\text{qm}}=t$, whatever be the Hamiltonian, its
time derivative operator is the unit operator.
     
     Let the classical Hamiltonian for a particle in four-dimensional
space-time be
\begin{equation}\label{E1:eq10}
H\ =\ \frac{1}{2m}(p_x^2+p_y^2+p_z^2)+V(t,x,y,z).
\end{equation}
Then \eqref{E1:eq5} can be solved for $p_z$ to yield
\begin{equation}\label{E1:eq11}
\mathcal{D}'_{\pm\text{cl}}(t,p_t,x,p_x,y,p_y,z,p_z)
\ =\ p_z\mp[-2m(p_t+V)-p_x^2-p_y^2]^{1/2}\ =\ 0.
\end{equation}
The quantity $\mathcal{D}'_{\pm\text{cl}}$, considered as a function
of eight variables prior to its role as a constraint,
generates the equations of motion for the evolution of the system in the
$\pm z$-direction.  The same classical dynamics is obtained if we return to
the original Lagrangian formalism, take $t$ as a dependent variable and 
$z$ as the independent variable, and procede to the Hamiltonian formalism.
We remark that the quantum operator $\hbar H_{\text{zev}}$ of \eqref{E2:eq11}
can be construed as a matrix square root of the quantum
operator that arises from a term in \eqref{E1:eq11}.

     In this paper we shall consider only the nonrelativistic
form of quantum mechanics, restricted to the problem of determining the
wave function of a massive, structureless particle in Galilean
four-dimensional space-time in the presence of a given space- and 
time-dependent potential energy distribution.  The physical
hypothesis that underlies the theory herein is that not just the
time component, but also the spatial components, of the 
conserved four-current density of probability flow are observables.
This hypothesis does not seem to have been made explicitly, or
its consequences studied, heretofore.  A sketch of the mathematics
introduced to elaborate this physical assumption follows.

     We shall treat quantum mechanics as a boundary-value problem 
for the Schr\"odinger equation.  The specified boundary values
on a simple domain in space-time
will be regarded as input to the problem, and the 
derived interior values 
and complementary boundary values will be considered as the 
overall output.  An intermediate objective will be to formulate
a theory of spatial evolution of wave functions.  
As is usual in applications, we shall
choose one spatial coordinate as the evolution coordinate,
and expand the wave function in terms of conveniently simple orthogonal
functions of the transverse coordinates, which include the time.
The time and the two transverse spatial coordinates
therefore will appear naturally as operator/observables
in the space of such functions, analogous to the role of
the three spatial positions when the time is taken as the
evolution coordinate.

    Suppose that we want to solve the Schr\"odinger equation in a space-time
box surrounded by two $t$=constant walls and a spatial
boundary.  Since the equation is
of first order in time and of second order in the spatial
coordinates, it is, mathematically speaking, necessary and sufficient to
supply wave function values on the earlier $t$=constant surface,
and a suitable combination of wave function and normal-derivative
values on the spatial boundary surface, to infer that an 
interior solution exists and is unique, 
as discussed in \cite{R:Friedman1}, Ch.\ 5, \S3.
(Our mathematics differs from Friedman's in that 
his equation is the heat equation rather than the
Schr\"odinger equation, and we shall
administer nonlocal boundary conditions,
which distinguish input from output signals, on the spatial boundary.)
Conventional time-dependent quantum mechanics for the
most part deals with specifying initial, or (but not and) occasionally final,
values on a $t$=constant surface and simple (often, zero) values
on the spatial boundary, which can be partly or wholly at infinity.
Nontrivial spatial boundary values,
as incoming wave amplitudes in a scattering problem, are conventionally
specified only in the context of the time-independent
Schr\"odinger equation.  In the present work we shall generalize
the latter problem by considering general time-dependent,
as well as space-dependent, input values on the
spatial boundary, in the presence of explicit time
dependence in the potential energy function in the differential
equation itself.  We shall hereinafter denote these
cases of boundary value problems as Type I and Type II, respectively.
These correspond, roughly and respectively,
to the first and second initial-boundary
value problems analyzed in \cite{R:Friedman1}, Chs.\ 3 and 5.

     In the first problem, the wave function evolves
in time from given initial values, with time-independent
spatial boundary values.  In the second problem, we shall
consider that the wave function evolves with respect to the $z$-coordinate,
such that the interior domain corresponds
to a finite interval in the chosen coordinate $z$.
Since the differential equation is of second order,
determining the evolution 
of a wave function in a spatial direction is generally a more difficult
task of analysis in both the mathematical and physical
senses than one for its evolution in time.  
We summarize the derivation to be carried out below
in terms of the following ten observations, steps, or results:
(i) the space of states on any given $z=z_1$
hyperplane has a natural doubled structure 
in that it comprises the direct sum of the values and of the 
$z$-derivatives of the usual space of wave functions $\psi(t,x,y,z)$
at $z=z_1$;
(ii) the Hamiltonian is a $2\times2$ matrix
of operators that is derived from the ordinary Schr\"odinger equation;
(iii) the familiar expression for the probability current density in the
$z$-direction is used to infer the definition of a
metric operator in the space of states, 
where now inner products include an integral over $t$
as well as over $x$ and $y$;
(iv) the metric so derived is indefinite, and the Hamiltonian
is self-adjoint 
with respect to the metric (synonymously, pseudo-Hermitean); 
(v) the norm being indefinite, we shall sometimes use the term
``particle presence'' to denote the
unit operator, the expectation value of which is the 
above-mentioned norm; 
(vi) apart from modifications needed for closed channels, 
the formalism can be established so that
waves traveling in the $+z$ direction
have positive norm, and waves traveling in the $-z$ direction have
negative norm with respect to the metric;
(vii) while for open-channel modes the direction
of travel and of propagation will coincide in the large,
it is convenient to define these categories differently
for closed-channel modes;
(viii) the input and output at either end of a finite spatial interval 
$[z_1,z_2]$ are taken
to comprise, respectively, the superposition of waves propagating into, and 
the superposition of waves propagating out of, the
interval at the initial point $z_1$ and at the final point $z_2$
(this means that there will be only outward 
propagating scattered waves from 
a zone of interaction);
(ix) orthonormal sets of input or of output states,
transition amplitudes, and probabilities
are then computed using what amounts
to a Hilbert space inner product, which is derived from the indefinite metric,
but depends on the wave function and its $z$-derivative
at both $z_1$ and $z_2$; (x) the dynamics yields a mapping of 
open-channel input into open-channel output that is unitary.  

     The fact that a pseudo-Hermitean Hamiltonian
describes the spatial evolution of a 
physical system's wave function has another
concomitant:  the Hamiltonian can have, as well as real eigenvalues,
nonreal eigenvalues that occur in complex conjugate pairs
(Gohberg, et al., \cite{R:Gohberg1} p. 23, Proposition 2.4).  Each such
pair is associated with the two wave function solutions 
(one rising, the other falling exponentially) for 
a closed channel, or classically inacessible region for the system when it
is in an associated quantum state.  We shall argue that it
is natural to define the direction of propagation 
(but not of travel of the particle that the wave represents---see 
the discussion in Sec.\ 4) of such a wave as
the direction in which it decreases exponentially.
The simple exponential states in such a pair each have zero norm and, with
proper normalization, unit overlap, which complicates the formalism.
A further complication results from the circumstance that
a degenerate eigenvalue of a Hamiltonian
requires special treatment when the Hamiltonian
cannot be diagonalized
by a similarity transformation, leading to the appearance of so-called
``$n$-pole ghost'' quantum states, for $n=2,3,\ldots$.
There is more discussion on these problems below.

     Formalisms for the spatial evolution of a wave function
were proposed by Kijowski \cite{R:Kijowski1}
and by Piron \cite{R:Piron1}, and their work was discussed by Mielnik
\cite{R:Mielnik1}.  These two approaches
differ substantially from 
each other and from the formalism introduced herein, as will
be discussed following Eq.\ \eqref{E2:eq11} and in Sec.\ 4.
 
     The quantum mechanics describing evolution of a wave function in both
directions across a spatial interval is to an extent
patterned after the author's previous work 
\cite{R:Hahne1} on a quantum dynamics that encompasses
joint bidirectional evolution of a quantum state between two
temporal walls.

     The remaining sections are organized as follows:  In Sec.\ 2,
we shall formulate expressions for the four-current density
associated with a physical quantity, and for the local
space-time density for creating or destroying that quantity
in a quantum-mechanical system.  We shall also show how
to prescribe physically motivated
boundary conditions so that the Schr\"odinger
equation can be solved in a semi-infinite 
(finite in the $z$-direction, infinite in the $t,x,y$-directions) box.  
In Section 3 we shall 
propose a formalism for computing the average temporal
position (i.e., crossing time) of the particle at both spatial walls
of the box, given the spatial input and given
the $S$-matrix deriving from a 
general interaction potential energy in the box's interior.
These results will then be used to compute formulas for dwell
and delay times for the particle remaining within, reflecting from, or 
transmitted across, the box.
Section 4 concludes the paper
with a discussion of the present formalism and
of previous work on the subject.  The matters of an uncertainty
principle involving the time and its conjugate momentum,
and of an extended theory of measurement, are
merely touched on there, as attempts by the author 
to develop these constructs have not been successful.


\section{Quantum-mechanical formalism} \label{S:sec2}

     In this section we shall set up the 
theory that forms the ``floor'' of the present work.
Rather than attempt to make the formalism highly general,
we shall develop the argument in a particular context:
the wave function solution of Schr\"odinger's equation 
for a particle moving in the interior of a certain simple box of 
four-dimensional space-time.  In particular, we propose a formalism 
and an interpretation that incorporate the wave function
into an expression for the space-time 
four-vector ``flow'' density of a physical
quantity, which quantity corresponds to a certain linear operator
in the function space of fully time- and space-dependent wave functions.
We shall argue that is is natural to regard the four-divergence of the
flow as the local density of creation and destruction of
that quantity at a point in space-time
for the physical system in that time-dependent quantum state.  Either the 
volume integral of the divergence, or the surface integral of the
normal component of the flow vector density, therefore
represents the total amount of that quantity generated
inside the space-time box.  If that quantity is the time $t$,
this integral plausibly represents the average so-called dwell time of
the particle in the given box, given that the wave function
is properly normalized.

     Let $\mathcal{B}_1$ and $\mathcal{B}_2$
be the following open boxes in space-time:
\begin{subequations}\label{E2:eq1}
\begin{align}
\mathcal{B}_1\,&=\,\{(t,x,y,z) |t_1<t<t_2, -\infty<x<\infty,
-\infty<y<\infty, -\infty<z<\infty\}.\label{E2:eq1a}\\
\mathcal{B}_2\,&=\,\{(t,x,y,z) |-\infty<t<\infty, -\infty<x<\infty,
-\infty<y<\infty, z_1<z<z_2\}.\label{E2:eq1b}
\end{align}
\end{subequations}
The Schr\"odinger equation for $\psi(t,x,y,z)$ for a particle of mass $m$
can be derived from a variational principle for an action 
$\mathcal{A}$, as given in Schiff (\cite{R:Schiff1}, p.\ 499), but
modified to make it real:
\begin{equation}\label{E2:eq2}
\mathcal{A}\ =\ \iiiint_{\mathcal{B}} dt\,dx\,dy\,dz\biggl[
\frac{i\hbar}{2}\psi^\ast\frac{\partial\psi}{\partial t}
-\frac{i\hbar}{2}\frac{\partial\psi^\ast}{\partial t}\psi\\
-\frac{\hbar^2}{2m}\mathbf{\nabla}\psi^\ast\cdot\mathbf{\nabla}\psi
-\psi^\ast V(t,x,y,z)\psi\biggr].
\end{equation}
The equations of motion are to be obtained by keeping the
boundary values of $\psi$ and $\psi^\ast$ fixed, and pretending that
in the interior region $\psi$ and $\psi^\ast$ can be varied
independently and ``arbitrarily''.   The action is stationary when
$\psi(t,x,y,z)$ satisfies
\begin{equation}\label{E2:eq3}
\frac{\hbar}{i}\frac{\partial\psi}{\partial t}
-\frac{\hbar^2}{2m}\nabla^2\psi+V(t,x,y,z)\psi\,=\,0,
\end{equation}
and $\psi(t,x,y,z)^\ast$ satisfies the complex conjugate equation,
for all $(t,x,y,z)\in\mathcal{B}_1\text{ or }\mathcal{B}_2$.

     Henceforth when we say ``solution'',
we shall mean a function defined over the entire box
such that it satisfies equation \eqref{E2:eq3} everywhere
in $\mathcal{B}_1$ or $\mathcal{B}_2$.  The linear operators
representing physical quantities will normally carry a
solution into another space- and time-dependent function
that is not a solution, so in effect we shall deal with the    
more general vector space of well-behaved, complex-valued functions
of space and time that need not be solutions
of the Schr\"odinger equation.

     A standard problem in conventional quantum mechanics
arises if we constrain $\psi(t,x,y,z)$ to be zero on 
the infinite parts of the spatial
boundary of the box $\mathcal{B}_1$, that is 
\begin{equation}\label{E2:eq4}
\psi(t,x,y,z)\to 0,\  
\text{if $|x|+|y|+|z|\to\infty$}.
\end{equation}
and require that
\begin{equation}\label{E2:eq5}
\psi(t,x,y,z)\to u(x,y,z),\ \ \text{as $t\to t_1$,}
\end{equation}
where $u(x,y,z)$ is some given complex-valued function on the 
earlier temporal boundary of the box.  As is well known, the interior
values of $\psi(t,x,y,z)$, and the limiting values on the
temporally later boundary of the box at $t=t_2$, are all uniquely
determined by the differential equation and these input boundary conditions.
The derived values are all output in a sense, but we shall often
mean by output just the subset of boundary values that were not
given as input, in the present case $\psi(t_2,x,y,z)$
with $-\infty<x<\infty$, $-\infty<y<\infty$, and $-\infty<z<\infty$.

    Now let us consider a problem 
in box $\mathcal{B}_2$ such that
certain information about the limiting values of $\psi(t,x,y,z)$
and $\partial\psi/\partial z(t,x,y,z)$ on the two walls
$z=z_1$ and $z=z_2$ is given as input, 
while the wave function is supposed to tend to zero
as $t$ and/or $x$ and/or $y$ tend to $\pm\infty$.
We want to specify just enough input information so that a solution
satisfying the input boundary conditions exists and is unique.
In order to accomplish this, we need to do some preliminary work.
We shall not keep to a mathematically rigorous derivation,
but appeal to plausibility arguments at most steps.

     We now convert the above variational principle to Hamiltonian
form using the methods of Goldstein (\cite{R:Goldstein1}, Chap.\ 12-4),
with the proviso that it is the spatial parameter $z$, rather than
$t$, that is taken as the evolution coordinate for the wave function.
The quantity in square brackets in \eqref{E2:eq2} is the Lagrangian
density $\mathcal{L}$.  The canonical field momenta are
\begin{subequations}
\begin{align}
p_{\psi}\ &=\ \frac{\partial\mathcal{L}}{\partial\Bigl(\frac{\partial\psi}
{\partial z}\Bigr)}
\ =\ -\frac{\hbar^2}{2m}\frac{\partial\psi^\ast}{\partial z},
\label{E2:eq6a}\\
p_{\psi^\ast}\ &=\ \frac{\partial\mathcal{L}}
{\partial\Bigl(\frac{\partial\psi^\ast}
{\partial z}\Bigr)}\ =\ -\frac{\hbar^2}{2m}\frac{\partial\psi}{\partial z}.
\label{E2:eq6b}
\end{align}
\end{subequations}
The action functional becomes
\begin{equation}\label{E2:eq7}
\mathcal{A}= \iiiint_{{\mathcal{B}}_2} dt\,dx\,dy\,dz\Bigl[p_\psi
\frac{\partial\psi}{\partial z}+\frac{\partial{\psi^\ast}}{\partial z}
p_{\psi^\ast}
-\mathcal{H}(\psi,p_\psi,\psi^\ast,p_{\psi^\ast})\Bigr],
\end{equation}
where the Hamiltonian density is
\begin{equation}\label{E2:eq8}
\mathcal{H}= -\frac{2m}{\hbar^2}p_\psi p_{\psi^\ast}
-\frac{i\hbar}{2}\psi^\ast\frac{\partial\psi}{\partial t}
+\frac{i\hbar}{2}\frac{\partial\psi^\ast}{\partial t}\psi
+\frac{\hbar^2}{2m}\Bigl(
\frac{\partial\psi^\ast}{\partial x}\frac{\partial\psi}{\partial x}+
\frac{\partial\psi^\ast}{\partial y}\frac{\partial\psi}{\partial y}
\Bigr)+\psi^\ast V\psi.
\end{equation}
The equations of motion obtained by varying $\psi^\ast$
and $p_\psi$ are a coupled set of linear equations; 
a complex conjugate set is obtained by varying $\psi$
and $p_{\psi^\ast}$.  We write the former equations in $2\times 2$
matrix-operator form as follows:  We first define
\begin{equation}\label{E2:eq9}
\Psi(t,x,y,z)\ =\ \left[\begin{matrix}\psi(t,x,y,z)\\
p_{\psi^\ast}(t,x,y,z)\end{matrix}\right];
\end{equation}
then the equations of motion can be written
\begin{equation}\label{E2:eq10}
\frac{1}{i}\frac{\partial\Psi}{\partial z}\ =\ H_{\text{zev}}\Psi,
\end{equation}
where the Hamiltonian $H_{\text{zev}}$ (the subscript ``zev'' stands
for ``$z$-evolution'') is
\begin{equation}\label{E2:eq11}
H_{\text{zev}}\ =\ \left[\begin{matrix} 0 &\frac{2mi}{\hbar^2}\\
\frac{1}{i}\Bigl[i\hbar\frac{\partial}{\partial t}
+\frac{\hbar^2}{2m}\Bigl(\frac{\partial^2}{\partial x^2}
+\frac{\partial^2}{\partial y^2}\Bigr)-V(t,x,y,z)\Bigr]&0\end{matrix}\right].
\end{equation}
     
     We note that Piron \cite{R:Piron1} 
obtained a Schr\"odinger equation for a wave function's
evolution along the spatial coordinate $x$, but Piron's wave function
has one component, and the Hamiltonian is the operator derived from
the classical quantity that generates dynamical motion along
a the $x$-axis.  Piron thereupon obtained a general expression for the
evolution in $x$ of the average temporal position of a particle
in one space dimension, but did not develop the theory further.

     In the ordinary quantum mechanics derivable from
the variational principle Eq.\ \eqref{E2:eq2}, the $z$-component
of the conserved probability four-current density is 
(\cite{R:Schiff1}, p.\ 27)
\begin{equation}\label{E2:eq12}
J_3(t,x,y,z)\ =\ \frac{\hbar}{2im}\Bigl(\psi^\ast
\frac{\partial\psi}{\partial z}-\frac{\partial\psi^\ast}{\partial z}\psi
\Bigl).
\end{equation}
In the present language this expression takes the form
\begin{equation}\label{E2:eq13}
J_3\ =\ \hbar^{-1}\Psi^\dagger M\Psi,
\end{equation}
where $M$ is the $2\times 2$ matrix
\begin{equation}\label{E2:eq14}
M\ =\ \left[\begin{matrix} 0&i\\-i&0\end{matrix}\right].
\end{equation}
Note that the matrix $M$ is Hermitean,
has unit square, and has eigenvalues $\pm 1$, so that it can
engender an indefinite metric.  
By inference, we make a guess for an inner product law
for two $z$-propagating states:
\begin{equation}\label{E2:eq15}
(\Psi_1(z);\Psi_2(z))\ =\ \hbar^{-1}
\iiint_{\mathbb{R}^3}dt\,dx\,dy\,
\bigl[\Psi_1(t,x,y,z)^\dagger\, M\Psi_2(t,x,y,z)\bigr].
\end{equation}
Note that this formula has the appropriate physical dimensions, in that
if $\psi_{1,2}$ have the usual dimension $\text{\it{length}}^{-3/2}$, then the
above inner product is dimensionless.

    We shall now argue that the above ingredients can be made into a
theory of spatial evolution of a Schr\"odinger wave function.
We shall work with the case of $z$-evolution of a wave function in
four-dimensional space-time, but generalizations to other cases,
as radial or reaction coordinates (see \cite{R:Miller1}) 
for the $(3N+1)$-dimensional
space-time involved in an $N$-particle wave function, are 
formally straighforward. 

     Let $\mathcal{S}$ be the space of functions of 
type Eq.\ \eqref{E2:eq9}, with some
appropriate boundary conditions.  
We define the $M$-adjoint of a linear operator 
$W$ acting on this space as that unique operator $W^\ddagger$ such that
\begin{equation}\label{E2:eq16}
(W^\ddagger\Psi_1;\Psi_2)\ =\ (\Psi_1;W\Psi_2)
\end{equation}
for all $\Psi_1,\Psi_2\in\mathcal{S}$;
in $2\times 2$ matrix form, with $W^\dagger$ as the
ordinary Hermitean conjugate, we have
\begin{equation}\label{E2:eq17}
W^\ddagger\ =\ MW^\dagger M.
\end{equation}
If 
\begin{equation}\label{E2:eq18}
W^\ddagger\ =\ W
\end{equation}
we call $W$ pseudo-Hermitean, and if
\begin{equation}\label{E2:eq19}
W^{\ddagger}\ =\ W^{-1}
\end{equation}
we call $W$ pseudo-unitary.

     If $W$ is pseudo-Hermitean and the state $\Psi$ is suitably normalized,
we want to make the plausible specification that
the (necessarily real) number $(\Psi;W\Psi)$ is the expectation
value of $W$ in the state $\Psi$.  We argue in favor
of this axiom as follows:  Let $x^0=t$, $x^1=x$, and so on.
Suppose that in the conventional Schr\"odinger formalism,
$\omega$ is some physical quantity, such as the time $\check{t}$,
the spatial positions $\check{x},\check{y},\check{z}$,
or the ``particle presence'' $\check{1}$
(we denote operators standing for physical
parameters with a ``ha\v{c}ek'' accent over the symbols).  
We take as a physical axiom that the four-vector ``flow'' density
$J_\mu^{(\omega)}(t,x,y,z)$, $\mu=0,1,2,3$, of $\omega$ is
\begin{subequations}\label{E2:eq20}
\begin{align}
J^{(\omega)}_0(t,x,y,z)
\ &=\ \psi(t,x,y,z)^\ast\omega\psi(t,x,y,z)\label{E2:eq20a}\\
J^{(\omega)}_k(t,x,y,z)\ &=\ \frac{\hbar}{2im}\biggl(
     \psi(t,x,y,z)^\ast\omega\frac{\partial\psi}{\partial x^k}(t,x,y,z)
\notag\\
&\ -\frac{\partial\psi^\ast}{\partial x^k}(t,x,y,z)\omega\psi(t,x,y,z)
\biggr),\ \text{for $k=1,2,3$.}\label{E2:eq20b}
\end{align}
\end{subequations}
The above expressions need symmetrization if $\omega$ contains
derivative operators, e.g., 
$\omega=X^j_{\text{cm}}=x^j+t(i\hbar/m)\partial/\partial x^j$,
one of the components of the initial center-of-mass position
for a free particle.
We compute the four-divergence of the above vector field, assuming that
$\psi$ is a solution to Eq.\ \eqref{E2:eq3}:
\begin{equation}\label{E2:eq21}
\begin{aligned}
\sum_{\mu=0}^3\frac{\partial J_\mu^{(\omega)}}{\partial x^\mu}(t,x,y,z)\ &=\ 
\frac{1}{i\hbar}\psi^\ast[\omega,V]\psi
+\psi^\ast\frac{\partial\omega}{\partial t}\psi\\
&+\frac{\hbar}{2im}\sum_{k=1}^3\biggl(
\psi^\ast\frac{\partial\omega}{\partial x^k}
\frac{\partial\psi}{\partial x^k}
-\frac{\partial\psi^\ast}{\partial x^k}
\frac{\partial\omega}{\partial x^k}\psi\biggr).
\end{aligned}
\end{equation}

     This divergence can be construed to be the local density
of creation or destruction of the quantity $\omega$ by the
system in the state $\psi(t,x,y,z)$.  If the divergence is zero,
as for the case $\omega=\check{1}$,
the associated quantity is not being
created or destroyed and is both globally and locally conserved.  
If $\omega=\check{t}$, we find that
\begin{equation}\label{E2:eq22}
\sum_{\mu=0}^3\frac{\partial J_\mu^{(\check{t})}}{\partial x^\mu}(t,x,y,z)\ =\ 
\psi^\ast\psi.
\end{equation}
Hence, the so-called ``dwell'' time $\tau_D$  of the
particle in a box $\mathcal{B}$ with the given input is
the space-time integral of the density of creation of time
over the box, that is,
\begin{equation}\label{E2:eq23}
\tau_D\ =\ \iiiint_{\mathcal{B}}dtdxdydz |\psi(t,x,y,z)|^2.
\end{equation}
The latter result reproduces a formula given in Ref.\ \cite{R:Chiao1}
Eq.\ (2.2), Ref.\ \cite{R:Muga2} Eq.\ (2.67), and
Ref.\ \cite{R:Nussenzweig1} Eq.\ (14).  By the divergence theorem
we can convert the volume integral to a surface integral,
so that we have either
\begin{subequations}\label{E2:eq24}
\begin{align}
\tau_D^{I}\ &=\ \biggl[\iiint_{\mathbb{R}^3}dx\,dy\,dz\,
J_0^{(\check{t})}(t,x,y,z)\biggr]\bigg|_{t=t_1}^{t=t_2},\text{\ or}
\label{E2:eq24a}\\
\tau_D^{II}\ &=\ 
\biggl[\iiint_{\mathbb{R}^3}dt\,dx\,dy\, J_3^{(\check{t})}(t,x,y,z)
\biggr]\bigg|^{z=z_2}_{z=z_1}.\label{E2:eq24b}
\end{align}
\end{subequations}
Hence if we have a boundary problem of Type I, such that
$\psi(t,x,y,z)$ is zero on the spatial walls, and $\psi$ 
has the usual conserved unit norm
on the $t$=constant walls, we find, with Eq.\ \eqref{E2:eq20a}
\begin{equation}\label{E2:eq25}
\tau_D^{I}\ =\ t_2-t_1.
\end{equation}
If we have a boundary value problem of Type II---we shall 
discuss later how to normalize $\psi$ in that case---so that 
$\psi(t,x,y,z)$ tends to zero as $|t|$ or $|x|$ or $|y|$ becomes large, 
and using Eq.\ \eqref{E2:eq20b} then
\begin{equation}\label{E2:eq26}
\begin{aligned}
\tau_D^{II}\ &=\ \biggl[\frac{\hbar}{2im}
\iiint_{\mathbb{R}^3}dt\,dx\,dy
\biggl(\psi(t,x,y,z)^\ast\,t\,\frac{\partial\psi}{\partial z}(t,x,y,z)\\
&\ -\frac{\partial\psi^\ast}{\partial z}(t,x,y,z)\,t\,\psi(t,x,y,z)
\biggr)\biggr]\bigg|^{z=z_2}_{z=z_1}.
\end{aligned}
\end{equation}
The alternate forms of the dwell time given in Eqs.\ \eqref{E2:eq23}
and \eqref{E2:eq26} were previously obtained by Jaworski and Wardlaw 
and applied in a series of papers (\cite{R:Jaworski1} Eqs.\ (4.2) and (A1),
\cite{R:Jaworski2}, \cite{R:Jaworski3}, \cite{R:Jaworski4}).
Given that we compute $\Psi(t,x,y,z)$
from $\psi(t,x,y,z)$ by Eq.\ \eqref{E2:eq9},
and that $I_2$ is the $2\times 2$ unit matrix, then the operator 
$\check{t}I_2$
is the time operator in the $z$-evolution formalism,
and we have, in the notation of Eq.\ \eqref{E2:eq15},
\begin{equation}\label{E2:eq27}
\tau_D^{II}\ =\ \bigl(\Psi(z);(\check{t}I_2)\Psi(z)\bigr)
\big|^{z=z_2}_{z=z_1}.
\end{equation}

     The above results suggest that for a pseudo-Hermitean
operator $W$ in the space of $\Psi$-solutions, 
and for a boundary value problem of Type II, we should
define the expectation value $\langle W\rangle_{\Psi(z)}$ of $W$ 
in the state $\Psi(t,x,y,z)$ at a chosen $z$ as
\begin{equation}\label{E2:eq28}
\langle W\rangle_{\Psi(z)}\ =\ 
\bigl(\Psi(z);W\Psi(z)\bigr),
\end{equation}
as was proposed earlier in this section.
The value $\langle W\rangle_{\Psi(z)}$
therefore (in Type II problems) specifies the average
net flow of $W$ across the given $z$=constant surface.
The difference of the expectation values of $W$ computed at
$z=z_2$ and $z=z_1$ is therefore 
the net flow of $W$ out of the box, in other words is,
on average, the total amount of $W$
``created'' by the system in the box.
 
     We want now to define input and output on the spatial walls
of the box.  
We define a complete, orthonormal basis $\phi_{(k_t,k_x,k_y)}(t,x,y)$
for all $(t,x,y)\in\mathbb{R}^3$ as follows:
\begin{equation}\label{E2:eq29}
\phi_{(k_t,k_x,k_y)}(t,x,y)\ =\ (2\pi)^{-3/2}\exp(-ik_tt+ik_xx+ik_yy),
\end{equation}
where $k_t$, $k_x$, and $k_y$ each range independently from 
$-\infty$ to $+\infty$.  Although the physical dimension of $k_t$
differs from that of $k_x$ and $k_y$, it is convenient to use three-vector
notation $\mathbf{k}=(k_t,k_x,k_y)$ and call the three-volume
element $d^3k=dk_tdk_xdk_y$.  The negative sign before $k_t$ in
the exponent in Eq.\ \eqref{E2:eq29} is chosen so that positive
$k_t$ corresponds to positive energy; the conjugate momentum
to $t$ is 
$p_t\leftrightarrow(\hbar/i)\partial/\partial t\leftrightarrow -\hbar k_t$.

     In an expansion of a wave function $\Psi(t,x,y,z)$ in the above basis
functions, we will encounter certain quantities repeatedly, so we
now define simplified notation for
them:  Let $\zeta$ take either value $F$ or $B$, 
which stand for forward and backward propagation along $z$, respectively.
We also take
\begin{equation}\label{E2:eq30}
\sigma(\zeta)\ =\ \begin{cases} +1&\text{if $\zeta=F$,}\\
-1&\text{if $\zeta=B$.}
\end{cases}
\end{equation}
If $(2mk_t/\hbar)>(k_x^2+k_y^2)$ (called an open channel), we define
\begin{equation}\label{E2:eq31}
k_z(\mathbf{k})\ =\ \bigl[2mk_t/\hbar-k_x^2-k_y^2\bigr]^{1/2},
\end{equation}
and if $(2mk_t/\hbar)<(k_x^2+k_y^2)$ (called a closed channel), we define
\begin{equation}\label{E2:eq32}
\kappa_z(\mathbf{k})\ =\ \bigl[-2mk_t/\hbar+k_x^2+k_y^2\bigr]^{1/2}.
\end{equation}
We shall normally just use $k_z$ and $\kappa_z$ without
explicitly citing their arguments, except that primed, double
primed, and triple primed arguments will be denoted,
respectively, by $k_z'$, $k_z''$,
and $k_z'''$, and similarly for $\kappa_z$.

     Let a wave function have the expansion in basis functions
\begin{equation}\label{E2:eq33}
\Psi(t,x,y,z)\ =\ \iiint_{\mathbb{R}^3}d^3k\sum_{\zeta=B}^F
f^\zeta(\mathbf{k})\phi_{\mathbf{k}}(t,x,y)
X^\zeta(\mathbf{k};z),
\end{equation}
where the $f^\zeta(\mathbf{k})$ are the expansion amplitudes, and
where the $X^\zeta(\mathbf{k};z)$ are normalized solutions for forward
or backward motion along $z$, which we construct as follows:
Substituting Eq.\ \eqref{E2:eq33} into Eq.\ \eqref{E2:eq10}, we find that
\begin{equation}\label{E2:eq34}
\frac{1}{i}\frac{dX^\zeta}{dz}(\mathbf{k};z)\ =\ H_{\text{zev}}(\mathbf{k};z)
X^\zeta(\mathbf{k};z),
\end{equation}
where, for $V$ a function of $z$ alone,
\begin{equation}\label{E2:eq35}
H_{\text{zev}}(\mathbf{k};z)\ =\ \left[\begin{matrix}
0&2mi/\hbar^2\\
(\hbar^2/2mi)\bigl[2mk_t/\hbar-k_x^2-k_y^2-2mV(z)/\hbar^2]&0
\end{matrix}\right].
\end{equation}
When $V(z)\equiv 0$, and for open channels, we obtain the solutions
\begin{equation}\label{E2:eq36}
X^\zeta(\mathbf{k};z)\ =\ \left[\begin{matrix}
[m/(\hbar k_z)]^{1/2}\exp[\sigma(\zeta)ik_zz]\\
-\sigma(\zeta)(i/2)(\hbar^3k_z/m)^{1/2}\exp[\sigma(\zeta)ik_zz]
\end{matrix}\right];\end{equation}
the corresponding inner products are independent of $z$:
\begin{equation}\label{E2:eq37}
\hbar^{-1}X^{\zeta'}(\mathbf{k};z)^\dagger M
X^\zeta(\mathbf{k};z)
\ =\ \delta^{\zeta'\zeta}\sigma(\zeta).
\end{equation}
Note, however, that these solutions do not satisfy the Cauchy
inequality, in that
$|\hbar^{-1}X^\zeta(\mathbf{k}',z)^\dagger M
X^\zeta(\mathbf{k},z)|=(1/2)(\sqrt{k_z'/k_z}+\sqrt{k_z/k_z'})$,
which is greater than $1$ unless $k_z'=k_z$.
For closed channels the solutions are
\begin{equation}\label{E2:eq38}
X^\zeta(\mathbf{k};z)\ =\ \left[\begin{matrix}
[m/(\hbar \kappa_z)]^{1/2}\exp[-\sigma(\zeta)(i\pi/4+\kappa_z z)]\\
\sigma(\zeta)(1/2)(\hbar^3\kappa_z/m)^{1/2}
\exp[-\sigma(\zeta)(i\pi/4+\kappa_z z)]
\end{matrix}\right];
\end{equation}
the inner products take the $z$-independent forms
\begin{equation}\label{E2:eq39}
\hbar^{-1}X^{\zeta'}(\mathbf{k};z)^\dagger M
X^\zeta(\mathbf{k};z)\ =\ 
\delta^{\zeta'F}\delta^{B\zeta} + \delta^{\zeta'B}\delta^{F\zeta}.
\end{equation}
 
    In general, the properties 
that distinguish between between the four types of state of motion
of a particle, 
that is open- versus closed-channel type, and $F$ versus $B$ type,
depend on the local behavior of the state vector
in wavenumber space $(k_t,k_x,k_y)$.
The corresponding position $(t,x,y)$ space forms of these
properties are nonlocal.  
As mentioned in Section 4, these properties are likely
to complicate an attempt to make a physical interpretation,
in the context of the
present formalism, of measurements at a given
$z$ of local properties in position $t$, $x$, or $y$.
This is in contrast to standard quantum mechanics with
$t$ as the evolution coordinate, where there is only
one type of state in $x,y,z$: $F$-type and open channel.

     We note that the intermediate free-particle case 
$2mk_t/\hbar=k_x^2+k_y^2$ gives rise to a ``dipole ghost'' state, in that
the reduced Hamiltonian on the rhs of Eq.\ \eqref{E2:eq35}
cannot be diagonalized by a similarity transformation.
The construct, which Heisenberg named (see \cite{R:Heisenberg1}, references
given therein, and \cite{R:Nagy1}, p.\ 14), derives from
the definition of the minimal polynomial of a 
finite-dimensional, square, complex matrix---see MacLane and Birkhoff
\cite{R:MacLane1}, Ch. IX.6:
Let $L$ be a pseudo-Hermitean operator such that there exists a
(real or nonreal) eigenvalue $\lambda$ of $L$, an integer $n\geq 2$,
and a state $X_{\lambda n}$
so that the state $(L-\lambda)^{n-1}X_{\lambda n}$ 
is not the zero state and is an eigenstate in that
$(L-\lambda)^nX_{\lambda n}=0$, 
then $X_{\lambda n}$ will be called an ``n-pole ghost'' state
of $L$ associated with the eigenvalue $\lambda$.
We presume that, for any given $L$,
and for each of its eigenvalues $\lambda$, there is a bounded
number---possibly zero---of types of ghosts associated with it.
The ghost states associated with fixed $L$ and $\lambda$,
and of different pole-orders $n$ and $m$, are linearly independent
of each other and of associated eigenstates;
the direct sum of all the eigenstates 
and of all the corresponding linearly independent ghost states is a complete
set of states in the overall space.

     Continuing with the zero-potential-energy,
intermediate-case solutions, we note that
the symbols F and B are not useful.
We take the solutions $X^\alpha(\hbar(k_x^2+k_y^2)/(2m),k_x,k_y;z)$
where $\alpha=1,2$, as follows:
\begin{subequations}\label{E2:eq40}
\begin{align}
X^1(\hbar(k_x^2+k_y^2)/(2m),k_x,k_y;z)\ &=\ \left[\begin{matrix}
[m\rho/\hbar]^{1/2}\\0 \end{matrix}\right],\label{E2:eq40a}\\
X^2(\hbar(k_x^2+k_y^2)/(2m),k_x,k_y;z)\ &=\ \left[\begin{matrix}
[m/(\rho\hbar)]^{1/2}2iz\\ -i[\hbar^3/(m\rho)]^{1/2}
\end{matrix}\right],\label{E2:eq40b}
\end{align}
\end{subequations}
where $\rho$ is an arbitrary positive number of dimension
\textit{length} introduced to make the components 
dimensionally consistent with Eqs.\ \eqref{E2:eq36} and \eqref{E2:eq38}.
The inner products are also $z$-independent:
\begin{equation}\label{E2:eq41}
\begin{aligned}
\hbar^{-1}X^{\alpha'}(\hbar(k_x^2+k_y^2)/(2m),k_x,k_y;z)^\dagger M&
X^{\alpha}(\hbar(k_x^2+k_y^2)/(2m),k_x,k_y;z)\\
&\ =\ \begin{cases} 0, &\text{if $\alpha'=\alpha$,}\\
                                +1, &\text{if $\alpha'\neq\alpha$.}
\end{cases}
\end{aligned}
\end{equation}
Physically, the states in the intermediate case propagate
parallel to any plane $z$=constant, that is, neither forward
nor backward along $z$.  
Note that the solution Eq.\ \eqref{E2:eq40a} is, and that of 
Eq.\ \eqref{E2:eq40b} is not, an eigenstate with eigenvalue zero
of the reduced Hamiltonian on the rhs of Eq.\ \eqref{E2:eq35};
in fact, the $X^2$ is a dipole ghost state
for any choice of $k_x$, $k_y$ and $z$.

    We next compute the inner product
at each $z$ of two free-particle
wave functions $\Psi(t,x,y,z)$ and $\Phi(t,x,y,z)$, when they have
expansion amplitudes $f^\zeta(\mathbf{k})$ and $g^\zeta(\mathbf{k})$,
respectively.  It is convenient to divide $\mathbf{k}$-space
into domains for open and closed channels:
\begin{subequations}\label{E2:eq42}
\begin{align}
\iiint_{\text{open}}d^3k\ &=\ 
\iint_{\mathbb{R}^2}dk_x dk_y
\int_{\hbar(k_x^2+k_y^2)/2m}^\infty dk_t,
\label{E2:eq42a}\\
\iiint_{\text{closed}}d^3k\ &=\ 
\iint_{\mathbb{R}^2}dk_x dk_y
\int^{\hbar(k_x^2+k_y^2)/2m}_{-\infty} dk_t,
\label{E2:eq42b}\\
\iiint_{\mathbb{R}^3}d^3k\ &=\ \iiint_{\text{open}}d^3k 
  +\iiint_{\text{closed}}d^3k.\label{E2:eq42c}
\end{align}
\end{subequations}
The inner product of $\Psi$ and $\Phi$
is $z$-independent, and takes the form
\begin{equation}\label{E2:eq43}
\begin{aligned}
(\Psi(z);\Phi(z))
\ &=\ \iiint_{\text{open}}d^3k\sum_{\zeta=B}^F\sigma(\zeta)
f^\zeta(\mathbf{k})^\ast g^\zeta(\mathbf{k})\\
&+\ \iiint_{\text{closed}}d^3k
\bigl[f^F(\mathbf{k})^\ast g^B(\mathbf{k})
+f^B(\mathbf{k})^\ast g^F(\mathbf{k})\bigr].
\end{aligned}
\end{equation}
Note that the subspace generated by ``$F$'' open-channel states 
has a positive definite norm, while the space
of ``$B$'' open-channel states has a negative definite norm;
each of these subspaces therefore
comprises a Hilbert space.  In the scattering phenomena analysed
in Section 3 we shall discover that the open-channel sub-matrix
of the $S$-matrix is unitary, and preserves the inner product
of two vectors belonging to a direct sum of these Hilbert spaces
referring to different $z$-planes,
assembled so that the sign of the inner product and metric are reversed
in the second subspace component---hence there is a positive definite
metric overall. The ``$F$'' and ``$B$'' 
Hilbert spaces on any $z$=constant plane are of limited utility,
as most linear operators encountered in the space of states
do not map such a Hilbert space into itself, but generate
superpositions of $F$- and $B$-states, and
of open- and closed-channel states. 

     The question of normalizing the space-evolving wave functions
can now be addressed:  If the potential $V(t,x,y,z)\neq 0$
in, and only in, the interior of the box $\mathcal{B}_2$,
a solution $\Psi(t,x,y,z)$ of Eq.\ \eqref{E2:eq10}
can be expressed in an expansion of the type Eq.\ \eqref{E2:eq33},
belonging to potential-free regions,
in the neighborhood of both $z=z_1$ and $z=z_2$, but
with different expansion amplitudes at each end
of the interval.  
We first define the basis functions
\begin{equation}\label{E2:eq44}
\Xi^\zeta(\mathbf{k};t,x,y,z)\ =\ (2\pi)^{-3/2}\exp(-ik_tt+ik_xx+ik_yy)
X^\zeta(\mathbf{k};z).
\end{equation}
We assume here and unless otherwise stated
that there is no closed-channel input, and adopt the following
conventions:
\begin{subequations}\label{E2:eq45}
\begin{align}
\Psi(t,x,y,z_1)\ &=\ \iiint_{\text{open}}d^3k\,
f^F_{\text{in}}(\mathbf{k})\Xi^F(\mathbf{k};t,x,y,z_1)\notag\\
&\ \ \ \ \ +\iiint_{\mathbb{R}^3}d^3k\,
f^B_{\text{out}}(\mathbf{k})\Xi^B(\mathbf{k};t,x,y,z_1),\label{E2:eq45a}\\
\Psi(t,x,y,z_2)\ &=\ \iiint_{\text{open}}d^3k\,
f^B_{\text{in}}(\mathbf{k})\Xi^B(\mathbf{k};t,x,y,z_2)\notag\\
&\ \ \ \ \ +\iiint_{\mathbb{R}^3}d^3k\,
f^F_{\text{out}}(\mathbf{k})\Xi^F(\mathbf{k};t,x,y,z_2),\label{E2:eq45b}
\end{align}
\end{subequations}
Since the flow of particle presence
is conserved, the norms of $\Psi(t,x,y,z_1)$
and $\Psi(t,x,y,z_2)$ are equal:
\begin{equation}\label{E2:eq46}
\langle \check{1}\rangle_{\Psi(z_2)}\ =\ 
\langle \check{1}\rangle_{\Psi(z_1)}.
\end{equation}
Note that at $z=z_1$, the forward and backward propagating parts
of the wave function correspond to input and output, respectively,
with the opposite association at $z=z_2$.
Note also that for a time-dependent potential energy there will
be scattering from open-channel input into both open- and
closed-channel output, which circumstance is accounted for in 
Eq.\ \eqref{E2:eq45}.   We now define a normalized, Type II
wave function as one for which the input amplitude function
is normalized to one, that is,
\begin{equation}\label{E2:eq47}
\begin{aligned}
1\ &=\ \iiint_{\text{open}}d^3k\,\bigl[|f^F_{\text{in}}(\mathbf{k})|^2
+|f^B_{\text{in}}(\mathbf{k})|^2\bigr]\\
&=\ \iiint_{\text{open}}d^3k\,\bigl[|f^F_{\text{out}}(\mathbf{k})|^2
+|f^B_{\text{out}}(\mathbf{k})|^2\bigr],
\end{aligned}
\end{equation}
where the second equation follows from Eqs.\ \eqref{E2:eq43},
\eqref{E2:eq45}, and \eqref{E2:eq46}.
Note that there is no contribution to the output normalization
from closed-channel amplitudes.

     We conclude the section by developing a formula for
the expectation value of the operator $\check{t}I_2$ in terms of the
wave-number space expansion amplitudes $f^\zeta(\mathbf{k})$. 
Let $\Psi(t,x,y,z)$ be as in Eq.\ \eqref{E2:eq33}.
Then we have
\begin{equation}\label{E2:eq48}
\begin{aligned}
\langle\check{t}I_2\rangle_{\Psi(z)}\ &=\ \hbar^{-1}
\iiint_{\mathbb{R}^3}dt\,dx\,dy\iiint_{\mathbb{R}^3}d^3k\sum_{\zeta=B}^F
\Psi(t,x,y,z)^{\dagger}M \\ 
&\ \ \times f^\zeta(\mathbf{k})
(2\pi)^{-3/2}t\exp[-ik_tt+ik_xx+ik_yy]X^\zeta(\mathbf{k};z).
\end{aligned}
\end{equation}
Replacing $t\exp[-ik_tt]$ by $i(\partial/\partial k_t)\exp[-ik_tt]$
and integrating by parts on $k_t$, we find that
\begin{equation}\label{E2:eq49}
\begin{aligned}
\langle\check{t}I_2\rangle_{\Psi(z)}\ &=\ \hbar^{-1}
\iiint_{\mathbb{R}^3}dt\,dx\,dy\iiint_{\mathbb{R}^3}d^3k\sum_{\zeta=B}^F
\Psi(t,x,y,z)^\dagger\,M\\
&\ \ \times(2\pi)^{-3/2}\exp[-ik_tt+ik_xx+ik_yy]
\frac{1}{i}\frac{\partial}{\partial k_t}
\Bigl[f^\zeta(\mathbf{k})X^\zeta(\mathbf{k};z)\Bigr].
\end{aligned}
\end{equation}
Analogous to spatial position operators in momentum space,
the operator for $t$ transforms into $-i\partial/\partial k_t$,
the sign difference being a result of the negative sign in the exponent
in Eq.\ \eqref{E2:eq29}.  This result agrees with that in 
Ref.\ \cite{R:Goldberger1}, Ch.\ 8, Eq.\ (286); see also
\cite{R:Kijowski1}, \S8.  Eq.\ \eqref{E2:eq49} can be interpreted
as yielding a quantum-mechanical value for the average arrival time,
or crossing time, of the particle 
in the state $\Psi$ at the given $z$=constant plane.
    
     The evaluation of Eq.\ \eqref{E2:eq49} is facilitated by the following
formulas:   for $\zeta=F$ and $B$ and for $\zeta'=B$ and $F$, respectively,
and for open channels,
\begin{subequations}\label{E2:eq50}
\begin{align}
\hbar^{-1}X^\zeta(\mathbf{k};z)^\dagger M\frac{1}{i}\frac{\partial X^\zeta}
{\partial k_t}(\mathbf{k};z)\ &=\ \frac{mz}{\hbar k_z},\\
\hbar^{-1}X^\zeta(\mathbf{k};z)^\dagger M\frac{1}{i}\frac{\partial X^{\zeta'}}
{\partial k_t}(\mathbf{k};z)
\ &=\ \sigma(\zeta)\frac{im}{2\hbar k_z^2}\exp(-\sigma(\zeta)2ik_zz),
\end{align}
\end{subequations}
while for closed channels
\begin{subequations}\label{E2:eq51}
\begin{align}
\hbar^{-1}X^\zeta(\mathbf{k};z)^\dagger M\frac{1}{i}\frac{\partial X^\zeta}
{\partial k_t}(\mathbf{k};z)\ &=\ -\sigma(\zeta)\frac{m}{2\hbar\kappa_z^2}
\exp(-\sigma(\zeta)2\kappa_z z),\\
\hbar^{-1}X^\zeta(\mathbf{k};z)^\dagger M\frac{1}{i}\frac{\partial X^{\zeta'}}
{\partial k_t}(\mathbf{k};z)
\ &=\ \sigma(\zeta)\frac{imz}{\hbar\kappa_z}.
\end{align}
\end{subequations}

     If we carry out the differentiations in the integrand of
Eq.\ \eqref{E2:eq49}, we find that
\begin{equation}\label{E2:eq52}
\begin{aligned}
&\langle\check{t}I_2\rangle_{\Psi(z)}\ =\ 
\iiint_{\text{open}}d^3k
\biggl[\sum_{\zeta=B}^F \Bigl(\sigma(\zeta)f^\zeta(\mathbf{k})^\ast
\frac{1}{i}\frac{\partial f^\zeta}{\partial k_t}(\mathbf{k})\Bigr)\\
&\ -\frac{im}{2\hbar k_z^2}\exp(+2ik_zz)f^B(\mathbf{k})^\ast f^F(\mathbf{k})\\
&\ +\frac{im}{2\hbar k_z^2}\exp(-2ik_zz)f^F(\mathbf{k})^\ast f^B(\mathbf{k})\\
&\ +\frac{mz}{\hbar k_z}\bigl(|f^F(\mathbf{k})|^2+|f^B(\mathbf{k})|^2\bigr)
\biggr]\ +\ \iiint_{\text{closed}}d^3k\\
&\times\biggl[
f^B(\mathbf{k})^\ast\frac{1}{i}\frac{\partial f^F}{\partial k_t}
(\mathbf{k})
+f^F(\mathbf{k})^\ast\frac{1}{i}\frac{\partial f^B}{\partial k_t}
(\mathbf{k})
\\ &-\frac{m}{2\hbar\kappa_z^2}\exp(-2\kappa_z z)|f^F(\mathbf{k})|^2
+\frac{m}{2\hbar\kappa_z^2}\exp(+2\kappa_z z)|f^B(\mathbf{k})|^2\\
&-\frac{imz}{\hbar\kappa_z}f^B(\mathbf{k})^\ast f^F(\mathbf{k})
+\frac{imz}{\hbar\kappa_z}f^F(\mathbf{k})^\ast f^B(\mathbf{k})\biggr].
\end{aligned}
\end{equation}
Due to the denominators involving $k_z^2$ or $\kappa_z^2$ in
the above, convergence of the integrals requires that
the $f^\zeta(\mathbf{k})$ approach zero sufficiently
rapidly as $\mathbf{k}$ approaches the boundary between open and
closed channels.  


\section{Scattering;  dwell and delay times}\label{S:sec3}

     In this section, we shall presume the presence of a generic
potential energy distribution $V(t,x,y,z)$, such that its support
is contained within the box $\mathcal{B}_2$.  The
potential energy gives rise to scattering of the 
(we presume, purely open-channel) input signals,
such that reflected and transmitted waves across the spectrum of 
$\mathbf{k}$, including both open and closed channels,
will comprise the output signal from the box.  We now assume that
our prescription for specifying the input yields necessary
and sufficient information such that a solution to the
Schr\"odinger equation within the box exists, satisfies the input
boundary conditions, and is unique.  Accordingly, the output
is determined by the input, and this association must be
linear in view of the linearity of the Schr\"odinger equation.
The linear operator specifying this association consists of
reflection and transmission coefficients, which can be assembled
into an $S$-matrix, which in turn---as we shall verify---has a submatrix,
referring to purely open-channel output as well as input, that is unitary.

     We presume that the
Schr\"odinger equation has been solved for all open-channel inputs,
and express the output linearly in terms of the input as follows:
\begin{equation}\label{E3:eq1}
\begin{aligned}
f^B_{\text{out}}(\mathbf{k})\ &=\ \iiint_{\text{open}}d^3k'\\
&\times\Bigl[
R^{BF}(\mathbf{k};\mathbf{k}')f^F_{\text{in}}(\mathbf{k}')\\
&+\ T^{BB}(\mathbf{k};\mathbf{k}')f^B_{\text{in}}(\mathbf{k}')\Bigr],
\end{aligned}
\end{equation}
\begin{equation}\label{E3:eq2}
\begin{aligned}
f^F_{\text{out}}(\mathbf{k})\ &=\ \iiint_{\text{open}}d^3k'\\
&\times\Bigl[
T^{FF}(\mathbf{k};\mathbf{k}')f^F_{\text{in}}(\mathbf{k}')\\
&+\ R^{FB}(\mathbf{k};\mathbf{k}')f^B_{\text{in}}(\mathbf{k}')\Bigr].
\end{aligned}
\end{equation}
The functions $T^{FF}$, $R^{FB}$, $R^{BF}$, and $T^{BB}$
are reflection and transmission coefficients, where the input-to-output
superscripts are to be read from right to left.  In Eqs.\ \eqref{E3:eq1}
and \eqref{E3:eq2}, 
consistent with Eqs.\ \eqref{E2:eq45a} and \eqref{E2:eq45b},
the reflection and transmission coefficients are
defined for the output parameter $k_t$ having either an open-
or a closed-channel value.

     For later convenience, we define
\begin{equation}\label{E3:eq3}
I^{\text{open}}(\mathbf{k}'-\mathbf{k}'')\ =\ 
\delta^{\text{open}}(k_t'-k_t'')\delta(k_x'-k_x'')\delta(k_y'-k_y''),
\end{equation}
where $\delta^{\text{open}}(k_t'-k_t'')$ is defined only for
both $k_t'$ and $k_t''$ corresponding to open channels.

     Let us now substitute Eqs.\ \eqref{E3:eq1} and \eqref{E3:eq2}
into Eq.\ \eqref{E2:eq47}.   We obtain a quadratic
expression in the input amplitudes $f^\zeta_{\text{in}}$
on both sides of the resulting equation.  Since these
amplitude functions are arbitrary, the coefficients of the four
quadratic terms must be equal.  We infer that, for both
$k_t'$ and $k_t''$ being of open-channel type,
\begin{equation}\label{E3:eq4}
\begin{aligned} 
&\iiint_{\text{open}}d^3k
\Bigl[T^{FF}(\mathbf{k};\mathbf{k}')^\ast
T^{FF}(\mathbf{k};\mathbf{k}'')\\
&+R^{BF}(\mathbf{k};\mathbf{k}')^\ast
R^{BF}(\mathbf{k};\mathbf{k}'')\Bigr]\\
&\ =\ I^{\text{open}}(\mathbf{k}'-\mathbf{k}''),
\end{aligned}
\end{equation}
\begin{equation}\label{E3:eq5}
\begin{aligned} 
&\iiint_{\text{open}}d^3k
\Bigl[R^{FB}(\mathbf{k};\mathbf{k}')^\ast
R^{FB}(\mathbf{k};\mathbf{k}'')\\
&+T^{BB}(\mathbf{k};\mathbf{k}')^\ast
T^{BB}(\mathbf{k};\mathbf{k}'')\Bigr]\\
&\ =\ I^{\text{open}}(\mathbf{k}'-\mathbf{k}''),
\end{aligned}
\end{equation}
\begin{equation}\label{E3:eq6}
\begin{aligned} 
&\iiint_{\text{open}}d^3k
\Bigl[R^{FB}(\mathbf{k};\mathbf{k}')^\ast
T^{FF}(\mathbf{k};\mathbf{k}'')\\
&+T^{BB}(\mathbf{k};\mathbf{k}')^\ast
R^{BF}(\mathbf{k};\mathbf{k}'')\Bigr]\ =\ 0,
\end{aligned}
\end{equation}
\begin{equation}\label{E3:eq7}
\begin{aligned} 
&\iiint_{\text{open}}d^3k
\Bigl[R^{BF}(\mathbf{k};\mathbf{k}')^\ast
T^{BB}(\mathbf{k};\mathbf{k}'')\\
&+T^{FF}(\mathbf{k};\mathbf{k}')^\ast
R^{FB}(\mathbf{k};\mathbf{k}'')\Bigr]\ =\ 0.
\end{aligned}
\end{equation}

     Let us now make up an $S$-matrix and its transpose conjugate
$S^\dagger$ from the reflection and transmission matrices.
In the following, the unprimed index $k_t$ ranges over all
real values, while $k_t'$ and $k_t''$ range over open-channel
values only:
\begin{equation}\label{E3:eq8}
S(\mathbf{k};\mathbf{k}'')
\ =\ \left[\begin{matrix}T^{FF}(\mathbf{k};\mathbf{k}'')&
R^{FB}(\mathbf{k};\mathbf{k}'')\\
R^{BF}(\mathbf{k};\mathbf{k}'')&
T^{BB}(\mathbf{k};\mathbf{k}'')\end{matrix}\right],
\end{equation}
\begin{equation}\label{E3:eq9}
S^\dagger(\mathbf{k}';\mathbf{k})
\ =\ \left[\begin{matrix}T^{FF}(\mathbf{k};\mathbf{k}')^\ast&
R^{BF}(\mathbf{k};\mathbf{k}')^\ast\\
R^{FB}(\mathbf{k};\mathbf{k}')^\ast&
T^{BB}(\mathbf{k};\mathbf{k}')^\ast\end{matrix}\right].
\end{equation}
It is convenient to define two submatrices of $S$, the open-channel
part $S_o$ and the closed-channel part $S_c$ as follows:
\begin{subequations}\label{E3:eq10}
\begin{align}
S_o(\mathbf{k};\mathbf{k}')\ &=\ S(\mathbf{k};\mathbf{k}'),
\ \ \text{for all $k_t>(\hbar/2m)(k_x^2+k_y^2)$,}\label{E3:eq10a}\\
S_c(\mathbf{k};\mathbf{k}')\ &=\ S(\mathbf{k};\mathbf{k}'),
\ \ \text{for all $k_t<(\hbar/2m)(k_x^2+k_y^2)$.}\label{E3:eq10b}
\end{align}
\end{subequations}
Then $S_o$ is unitary on the left as a result of
Eqs.\ \eqref{E3:eq4}--\eqref{E3:eq7}:
\begin{equation}\label{E3:eq11}
\begin{aligned}
(S_o^\dagger S_o)(\mathbf{k}';\mathbf{k}'')
&=\ \iiint_{\text{open}}d^3k
\bigl[ S^\dagger_o(\mathbf{k}';\mathbf{k})
S_o(\mathbf{k};\mathbf{k}'')\bigr]\\
&=\ I_2\otimes I^{\text{open}}(\mathbf{k}'-\mathbf{k}'').
\end{aligned}
\end{equation}
One expects that $S_o$ is also  unitary on the right, 
\begin{equation}\label{E3:eq12}
(S_oS_o^\dagger)(\mathbf{k}';\mathbf{k}'')
=I_2\otimes I^{\text{open}}(\mathbf{k}'-\mathbf{k}'').
\end{equation}

     We now reduce the formulas for
the expectation values of the operator $\check{t}I_2$ at $z=z_1,z_2$,
using Eqs.\ \eqref{E2:eq45}, \eqref{E2:eq52}, 
\eqref{E3:eq1}, and \eqref{E3:eq2},
and then establish a relatively simple form for the difference
of the two values.  
We assume that the 
input amplitudes $f^\zeta_{\text{in}}(\mathbf{k})$ and the output state
values of the $S$-matrix elements go to zero
at the open/closed-channel threshold so that the following integrals converge.
We have first
\begin{equation}\label{E3:eq13}
\begin{aligned}
\langle \check{t}I_2
\rangle_{\Psi(z_1)} &=\ \iiint_{\text{open}}
d^3k\,f^F_{\text{in}}(\mathbf{k})^\ast
\frac{1}{i}\frac{\partial f^F_{\text{in}}}{\partial k_t}
(\mathbf{k})\\
&+\iiint_{\text{open}}d^3k\iiint_{\text{open}} 
d^3k'\sum_{\zeta,\zeta'}\\
&\times f^\zeta_{\text{in}}(\mathbf{k})^\ast A^{\zeta\zeta'}
(\mathbf{k};\mathbf{k}';z_1)
f^{\zeta'}_{\text{in}}(\mathbf{k}').
\end{aligned}
\end{equation}
The matrix of coefficients is as follows:
\begin{equation}\label{E3:eq14}
\begin{aligned}
A^{FF}(&\mathbf{k};\mathbf{k}';z_1) \\
&=\ \frac{mz_1}{\hbar k_z}I^{\text{open}}(\mathbf{k}-\mathbf{k}')\\
&+\frac{im}{2\hbar k^2_z}\exp(-2ik_z z_1)R^{BF}(\mathbf{k};\mathbf{k}')\\
&-R^{BF}(\mathbf{k}';\mathbf{k})^\ast
\frac{im}{2\hbar k_z'^2 }\exp(+2ik_z' z_1)\\
&+\iiint_{\text{open}}d^3k''\,
R^{BF}(\mathbf{k}'';\mathbf{k})^\ast\\
&\times\biggl[\frac{mz_1}{\hbar k_z''}
-\frac{1}{i}\frac{\partial}{\partial k_t''}\biggr]
 R^{BF}(\mathbf{k}'';\mathbf{k}')\\
&+\iiint_{\text{closed}}d^3k''
\,R^{BF}(\mathbf{k}'';\mathbf{k})^\ast\\
&\times\frac{m}{2\hbar\kappa_z''^2}\exp(2\kappa_z'' z_1)
R^{BF}(\mathbf{k}'';\mathbf{k}');
\end{aligned}
\end{equation}
\begin{equation}\label{E3:eq15}
\begin{aligned}
A^{FB}(&\mathbf{k};\mathbf{k}';z_1) \\
&=\ \frac{im}{2\hbar k_z^2}\exp(-2ik_z z_1)
T^{BB}(\mathbf{k};\mathbf{k}')\\
&+\iiint_{\text{open}}d^3k''
\,R^{BF}(\mathbf{k}'';\mathbf{k})^\ast\\
&\times\biggl[\frac{mz_1}{\hbar k_z''}
-\frac{1}{i}\frac{\partial}{\partial k_t''}\biggr]
T^{BB}(\mathbf{k}'';\mathbf{k}')\\
&+\iiint_{\text{closed}}d^3k''
\,R^{BF}(\mathbf{k}'';\mathbf{k})^\ast\\
&\times\frac{m}{2\hbar\kappa_z''^2}\exp(2\kappa_z'' z_1)
T^{BB}(\mathbf{k}'';\mathbf{k}');
\end{aligned}
\end{equation}
\begin{equation}\label{E3:eq16}
\begin{aligned}
A^{BF}(&\mathbf{k};\mathbf{k}';z_1) \\
&=\ -T^{BB}(\mathbf{k}';\mathbf{k})^\ast
\frac{im}{2\hbar k_z'^2}\exp(2ik'_z z_1)\\
&+\iiint_{\text{open}}d^3k''
\,T^{BB}(\mathbf{k}'';\mathbf{k})^\ast\\
&\times\biggl[\frac{mz_1}{\hbar k_z''}
-\frac{1}{i}\frac{\partial}{\partial k_t''}\biggr]
R^{BF}(\mathbf{k}'';\mathbf{k}')\\
&+\iiint_{\text{closed}}d^3k''
\,T^{BB}(\mathbf{k}'';\mathbf{k})^\ast\\
&\times\frac{m}{2\hbar\kappa_z''^2}\exp(2\kappa_z'' z_1)
 R^{BF}(\mathbf{k}'';\mathbf{k}');
\end{aligned}
\end{equation}
\begin{equation}\label{E3:eq17}
\begin{aligned}
A^{BB}(&\mathbf{k};\mathbf{k}';z_1) \\
&=\iiint_{\text{open}}d^3k''
\,T^{BB}(\mathbf{k}'';\mathbf{k})^\ast\\
&\times\biggl[\frac{mz_1}{\hbar k_z''}
-\frac{1}{i}\frac{\partial}{\partial k_t''}\biggr]
T^{BB}(\mathbf{k}'';\mathbf{k}')\\
&+\iiint_{\text{closed}}d^3k''
\,T^{BB}(\mathbf{k}'';\mathbf{k})^\ast\\
&\times\frac{m}{2\hbar\kappa_z''^2}\exp(2\kappa_z'' z_1)
T^{BB}(\mathbf{k}'';\mathbf{k}').
\end{aligned}
\end{equation}
At $z=z_2$ we have
\begin{equation}\label{E3:eq18}
\begin{aligned}
\langle \check{t}I_2\rangle_{\Psi(z_2)} &=\ -\iiint_{\text{open}}
d^3k\,f^B_{\text{in}}(\mathbf{k})^\ast
\frac{1}{i}\frac{\partial f^B_{\text{in}}}{\partial k_t}
(\mathbf{k})\\
&+\iint_{\text{open}}d^3k\iiint_{\text{open}} 
d^3k'\sum_{\zeta,\zeta'}\\
&\times f^\zeta_{\text{in}}(\mathbf{k})^\ast C^{\zeta\zeta'}
(\mathbf{k};\mathbf{k}';z_2)
f^{\zeta'}_{\text{in}}(\mathbf{k}').
\end{aligned}
\end{equation}
In the above, the coefficient matrices are
\begin{equation}\label{E3:eq19}
\begin{aligned}
C^{FF}(&\mathbf{k};\mathbf{k}';z_2) \\
&=\iiint_{\text{open}}d^3k''
\,T^{FF}(\mathbf{k}'';\mathbf{k})^\ast\\
&\times\biggl[\frac{mz_2}{\hbar k_z''}
+\frac{1}{i}\frac{\partial}{\partial k_t''}\biggr]
 T^{FF}(\mathbf{k}'';\mathbf{k}')\\
&-\iiint_{\text{closed}}d^3k''
\,T^{FF}(\mathbf{k}'';\mathbf{k})^\ast\\
&\times\frac{m}{2\hbar\kappa_z''^2}\exp(-2\kappa_z'' z_2)
 T^{FF}(\mathbf{k}'';\mathbf{k}');
\end{aligned}
\end{equation}
\begin{equation}\label{E3:eq20}
\begin{aligned}
C^{FB}(&\mathbf{k};\mathbf{k}';z_2) \\
&=\ T^{FF}(\mathbf{k}';\mathbf{k})^\ast
\frac{im}{2\hbar k_z'^2}\exp(-2ik'_z z_2)\\
&+\iiint_{\text{open}}d^3k''
\,T^{FF}(\mathbf{k}'';\mathbf{k})^\ast\\
&\times\biggl[\frac{mz_2}{\hbar k_z''}
+\frac{1}{i}\frac{\partial}{\partial k_t''}\biggr]
R^{FB}(\mathbf{k}'';\mathbf{k})\\
&-\iiint_{\text{closed}}d^3k''
\,T^{FF}(\mathbf{k}'';\mathbf{k})^\ast\\
&\times\frac{m}{2\hbar\kappa_z''^2}\exp(-2\kappa_z'' z_2)
R^{FB}(\mathbf{k}'';\mathbf{k}');
\end{aligned}
\end{equation}
\begin{equation}\label{E3:eq21}
\begin{aligned}
C^{BF}(&\mathbf{k};\mathbf{k}';z_2) \\
&=\ -\frac{im}{2\hbar k_z^2}\exp(2ik_z z_2)
T^{FF}(\mathbf{k};\mathbf{k}')\\
&+\iiint_{\text{open}}d^3k''
\,R^{FB}(\mathbf{k}'';\mathbf{k})^\ast\\
&\times\biggl[\frac{mz_2}{\hbar k_z''}
+\frac{1}{i}\frac{\partial}{\partial k_t''}\biggr]
T^{FF}(\mathbf{k}'';\mathbf{k}')\\
&-\iiint_{\text{closed}}d^3k''
\,R^{FB}(\mathbf{k}'';\mathbf{k})^\ast\\
&\times\frac{m}{2\hbar\kappa_z''^2}\exp(-2\kappa_z'' z_2)
T^{FF}(\mathbf{k}'';\mathbf{k}');
\end{aligned}
\end{equation}
\begin{equation}\label{E3:eq22}
\begin{aligned}
C^{BB}(&\mathbf{k};\mathbf{k}';z_2) \\
&=\ \frac{mz_2}{\hbar k_z}I^{\text{open}}(\mathbf{k}-\mathbf{k}')\\
&-\frac{im}{2\hbar k^2_z}\exp(2ik_z z_2)R^{FB}(\mathbf{k};\mathbf{k}')\\
&+R^{FB}(\mathbf{k}';\mathbf{k})^\ast
\frac{im}{2\hbar k_z' }\exp(-2ik_z' z_2)\\
&+\iiint_{\text{open}}d^3k''
\,R^{FB}(\mathbf{k}'';\mathbf{k})^\ast\\
&\times\biggl[\frac{mz_2}{\hbar k_z''}
+\frac{1}{i}\frac{\partial}{\partial k_t''}\biggr]
 R^{FB}(\mathbf{k}'';\mathbf{k}')\\
&-\iiint_{\text{closed}}d^3k''
\,R^{FB}(\mathbf{k}'';\mathbf{k}')^\ast\\
&\times\frac{m}{2\hbar\kappa_z''^2}\exp(-2\kappa_z'' z_2)
R^{FB}(\mathbf{k}'';\mathbf{k}').
\end{aligned}
\end{equation}

    The difference of the two expectation values takes the form
\begin{equation}\label{E3:eq23}
\begin{aligned}
\langle \check{t}I_2\rangle_{\Psi(z_2)}
-&\langle \check{t}I_2\rangle_{\Psi(z_1)}
\ =\ \iiint_{\text{open}}d^3k\iiint_{\text{open}} 
d^3k'\\ &\times\sum_{\zeta,\zeta'}
 f^\zeta_{\text{in}}(\mathbf{k})^\ast D^{\zeta\zeta'}
(\mathbf{k};\mathbf{k}';z_1;z_2)
f^{\zeta'}_{\text{in}}(\mathbf{k}').
\end{aligned}
\end{equation}
We break the $D$-matrices into constituents:
\begin{equation}\label{E3:eq24}
D\ =\ D_1+\,MD_2S_o+S_o^\dagger D_2^\dagger M
+S_o^\dagger D_3S_o+S_c^\dagger D_4S_c,
\end{equation}
where
\begin{equation}\label{E3:eq25}
D_1\ =\ I^{\text{open}}(\mathbf{k}-\mathbf{k}')\otimes
\text{diag}\biggl(
-\frac{1}{i}\frac{\partial}{\partial k_t'}
-\frac{mz_1}{\hbar k_z'},
-\frac{1}{i}\frac{\partial}{\partial k_t'}
+\frac{mz_2}{\hbar k_z'}\biggr),
\end{equation}
\begin{equation}\label{E3:eq26}
D_2\ =\ \text{diag}\Biggl(\frac{m}{2\hbar k_z^2}\exp(2ik_zz_2),
-\frac{m}{2\hbar k_z^2}\exp(-2ik_zz_1)\biggr),
\end{equation}
\begin{equation}\label{E3:eq27}
D_3\ =\ \text{diag}\biggl(\frac{1}{i}\frac{\partial}{\partial k_t''}
+\frac{mz_2}{\hbar k_z''},
\frac{1}{i}\frac{\partial}{\partial k_t''}
-\frac{mz_1}{\hbar k_z''}\biggr)
\end{equation}
\begin{equation}\label{E3:eq28}
D_4\ =\ \text{diag}\biggl(
-\frac{m}{2\hbar\kappa_z''^2}\exp(-2\kappa_z'' z_2),
-\frac{m}{2\hbar\kappa_z''^2}\exp(2\kappa_z'' z_1)\biggr).
\end{equation}
In Eqs.\ \eqref{E3:eq27} and \eqref{E3:eq28}, the double primes
indicate the dummy variables of integration implicit in
the final two summands on the rhs of Eq.\ \eqref{E3:eq24}.
The $D_1$ and $D_3$ terms were obtained by Smith \cite{R:Smith1}, 
\cite{R:Smith2} for the
case that the evolution coordinate is radial and the interaction
is time-independent, but otherwise general.  Note that Smith's
$S$-matrix is the transpose of the $S$-matrix defined above. 

     It is plausible that Eq.\ \eqref{E3:eq23},
given that the total input is normalized
as in Eq.\ \eqref{E2:eq47}, provides a 
complete expression for the mean dwell time of the particle in the box,
inasmuch as it is also an expression for the space-time integral
over the box of the divergence of the flow vector density of time.
We remark that if the potential energy is time-independent, then
the $S_o$-matrix takes the form $S_o(\mathbf{k}'';\mathbf{k}')
=\delta^{\text{open}}(k_t''-k_t')\bar{S}_o(k_t'',k_x'',k_y'';k_x',k_y')$;
one can now show that, due to the unitarity of $S_o$,
the terms involving 
$-i\partial f^\zeta_{\text{in}}/\partial k_t'(\mathbf{k}')$
cancel out in the overall expression for the dwell time.
This cancellation does not, as we shall see, occur for the
individual delay times for transmission or 
reflection from a zone of time-independent interaction.

     The average delay times that are measured in beam experiments for either
transmission or reflection are not so fundamentally defined.
We simplify the problem as follows:
First, we assume that only one kind
of input, that is F or but not and B, is present.
Second, we assume that the closed-channel contributions will be negligible
in the measuring apparatus.  Third, we neglect interference
between the incoming signal and the outgoing signal
in the case of reflection (hence, the contributions linear in the
$S$-matrix are discarded).  We now specify what remains after
these simplifications.

    In the first instance, let $f^B_{\text{in}}(\mathbf{k})\equiv 0$, and
let $f^F_{\text{in}}$ be normalized as in Eq.\ \eqref{E2:eq47}.
The net outgoing reflected and transmitted 
currents of particle presence are called 
$\mathcal{R}_{B\leftarrow F}^{\text{out}}(z_1)$ 
and $\mathcal{T}_{F\leftarrow F}^{\text{out}}(z_2)$, 
respectively, and take the values
\begin{subequations}\label{E3:eq29}
\begin{align}
\mathcal{R}_{B\leftarrow F}^{\text{out}}(z_1)
\ &=\ \iiint_{\text{open}}d^3k\iiint_{\text{open}}d^3k'
\iiint_{\text{open}}d^3k''\bigl[
f^F_{\text{in}}(\mathbf{k})^\ast \notag\\&\ \ \times
(-1)R^{BF}(\mathbf{k}'';\mathbf{k})^\ast
 R^{BF}(\mathbf{k}'';\mathbf{k}')
f^F_{\text{in}}(\mathbf{k}')\bigr],\label{E3:eq29a}\\
\mathcal{T}_{F\leftarrow F}^{\text{out}}(z_2)
\ &=\ \iiint_{\text{open}}d^3k\iiint_{\text{open}}d^3k'
\iiint_{\text{open}}d^3k''\bigl[
f^F_{\text{in}}(\mathbf{k})^\ast \notag\\
&\ \ \times T^{FF}(\mathbf{k}'';\mathbf{k})^\ast
T^{FF}(\mathbf{k}'';\mathbf{k}')
f^F_{\text{in}}(\mathbf{k}')\bigr].\label{E3:eq29b}
\end{align}
\end{subequations}
According to Eq.\ \eqref{E3:eq4}, we have
\begin{equation}\label{E3:eq30}
\mathcal{T}_{F\leftarrow F}^{\text{out}}(z_2)
\,-\,\mathcal{R}_{B\leftarrow F}^{\text{out}}(z_1)\ =\ 1.
\end{equation}
The mean currents of time at entry and upon reflection at $z_1$, 
and upon transmission at $z_2$, will be called, respectively,
$\tau_{F}^{\text{in}}(z_1)$, 
$\tau_{B\leftarrow F}^{\text{out}}(z_1)$, 
and $\tau_{F\leftarrow F}^{\text{out}}(z_2)$, 
and can be inferred from Eqs.\ \eqref{E3:eq14} (twice) and \eqref{E3:eq19},
subject to the three simplifications spelled out in the
previous paragraph, as follows:
\begin{subequations}\label{E3:eq31}
\begin{align}
\tau_{F}^{\text{in}}(z_1)\ &=\ \iiint_{\text{open}}d^3k
f^F_{\text{in}}(\mathbf{k})^\ast\biggl[\frac{1}{i}
\frac{\partial}{\partial k_t}\,+\,\frac{mz_1}{\hbar k_z}\biggr]
f^{F}_{\text{in}}(\mathbf{k}),\label{E3:eq31a}\\
\tau_{B\leftarrow F}^{\text{out}}(z_1)\ &=\ 
\iiint_{\text{open}}d^3k\iiint_{\text{open}}d^3k'
\biggl\{f^F_{\text{in}}(\mathbf{k})^\ast
\iiint_{\text{open}}d^3k'' R^{BF}(\mathbf{k}'';\mathbf{k})^\ast
\notag\\
&\times\biggl[-\frac{1}{i}\frac{\partial}{\partial k_t''}\,+\,
\frac{mz_1}{\hbar k_z''}\biggr]R^{BF}(\mathbf{k}'';\mathbf{k'})
\biggr]f^F_{\text{in}}(\mathbf{k}')\biggr\},\label{E3:eq31b}\\
\tau_{F\leftarrow F}^{\text{out}}(z_2)\ &=\ 
\iiint_{\text{open}}d^3k\iiint_{\text{open}}d^3k'\biggl\{
f^F_{\text{in}}(\mathbf{k})^\ast
\iiint_{\text{open}}d^3k'' T^{FF}(\mathbf{k}'';\mathbf{k})^\ast
\notag\\
&\times\biggl[\frac{1}{i}\frac{\partial}{\partial k_t''}\,+\,
\frac{mz_2}{\hbar k_z''}\biggr]T^{FF}(\mathbf{k}'';\mathbf{k'})
\biggr]f^F_{\text{in}}(\mathbf{k}')\biggr\}.\label{E3:eq31c}
\end{align}
\end{subequations}

    In the second instance, let $f^F_{\text{in}}(\mathbf{k})\equiv 0$, and
let $f^B_{\text{in}}(\mathbf{k})$ be normalized as in Eq.\ \eqref{E2:eq47}.
The net outgoing reflected and transmitted currents 
of particle presence are called 
$\mathcal{R}_{F\leftarrow B}^{\text{out}}(z_2)$ 
and $\mathcal{T}_{B\leftarrow B}^{\text{out}}(z_1)$, 
respectively, and take the values
\begin{subequations}\label{E3:eq32}
\begin{align}
\mathcal{R}_{F\leftarrow B}^{\text{out}}(z_2)
\ &=\ \iiint_{\text{open}}d^3k\iiint_{\text{open}}d^3k'
\iiint_{\text{open}}d^3k''\bigl[
f^B_{\text{in}}(\mathbf{k})^\ast 
\notag\\&\ \ \times R^{FB}(\mathbf{k}'';\mathbf{k})^\ast
R^{FB}(\mathbf{k}'';\mathbf{k}')
f^B_{\text{in}}(\mathbf{k}')\bigr],\label{E3:eq32a}\\
\mathcal{T}_{B\leftarrow B}^{\text{out}}(z_1)
\ &=\ \iiint_{\text{open}}d^3k\iiint_{\text{open}}d^3k'
\iiint_{\text{open}}d^3k''\bigl[
f^B_{\text{in}}(\mathbf{k})^\ast T^{BB}(\mathbf{k}'';\mathbf{k})^\ast
\notag\\
&\ \ \times (-1) T^{BB}(\mathbf{k}'';\mathbf{k}')
f^B_{\text{in}}(\mathbf{k}')\bigr].\label{E3:eq32b}
\end{align}
\end{subequations}
According to Eq.\ \eqref{E3:eq5}, we have
\begin{equation}\label{E3:eq33}
\mathcal{R}_{F\leftarrow B}^{\text{out}}(z_2)
\,-\,\mathcal{T}_{B\leftarrow B}^{\text{out}}(z_1)\ =\ 1.
\end{equation}
The net currents of time at $z_2$ upon entry and after reflection,
and at $z_1$ after transmission,
will be called, respectively,
$\tau_{B}^{\text{in}}(z_2)$, 
$\tau_{F\leftarrow B}^{\text{out}}(z_2)$, 
and $\tau_{B\leftarrow B}^{\text{out}}(z_1)$, 
and can be inferred from Eqs.\ \eqref{E3:eq22} (twice) and \eqref{E3:eq17},
subject to the three simplifications given previously, as follows:
\begin{subequations}\label{E3:eq34}
\begin{align}
\tau_{B}^{\text{in}}(z_2)\ &=\ \iiint_{\text{open}}d^3k
f^B_{\text{in}}(\mathbf{k})^\ast\biggl[-\frac{1}{i}
\frac{\partial}{\partial k_t}\,+\,\frac{mz_2}{\hbar k_z}\biggr]
f^{B}_{\text{in}}(\mathbf{k}),\label{E3:eq34a}\\
\tau_{F\leftarrow B}^{\text{out}}(z_2)\ &=\ 
\iiint_{\text{open}}d^3k\iiint_{\text{open}}d^3k'
\biggl\{f^B_{\text{in}}(\mathbf{k})^\ast
\iiint_{\text{open}}d^3k'' R^{FB}(\mathbf{k}'';\mathbf{k})^\ast
\notag\\
&\times\biggl[\frac{1}{i}\frac{\partial}{\partial k_t''}\,+\,
\frac{mz_2}{\hbar k_z''}\biggr]R^{FB}(\mathbf{k}'';\mathbf{k'})
\biggr]f^B_{\text{in}}(\mathbf{k}')\biggr\},\label{E3:eq34b}\\
\tau_{B\leftarrow B}^{\text{out}}(z_1)\ &=\ 
\iiint_{\text{open}}d^3k\iiint_{\text{open}}d^3k'\biggl\{
f^B_{\text{in}}(\mathbf{k})^\ast
\iiint_{\text{open}}d^3k'' T^{BB}(\mathbf{k}'';\mathbf{k})^\ast
\notag\\
&\times\biggl[-\frac{1}{i}\frac{\partial}{\partial k_t''}\,+\,
\frac{mz_1}{\hbar k_z''}\biggr]T^{BB}(\mathbf{k}'';\mathbf{k'})
\biggr]f^B_{\text{in}}(\mathbf{k}')\biggr\}.\label{E3:eq34c}
\end{align}
\end{subequations}

     We now undertake to use the derived results to obtain estimates for the
average delay time for the four processes of transmission and reflection.
Due to the absence of space- and time-reversal symmetry of the potential
energy, there will be no special relationships
between the two transmission times or between the two
reflection times.   Let the transmission delay times be called
$\tau^{\text{trans}}_{F\leftarrow F}(z_2\leftarrow z_1)$ and
$\tau^{\text{trans}}_{B\leftarrow B}(z_1\leftarrow z_2)$,
while the reflection delay times are called
$\tau^{\text{refl}}_{B\leftarrow F}(z_1\leftarrow z_1)$ and
$\tau^{\text{refl}}_{F\leftarrow B}(z_2\leftarrow z_2)$.
We proceed from the following principle for computing delay times
(currents are taken with their algebraic signs intact):
\begin{equation}\label{E3:eq35}
\begin{aligned}
\text{delay time}\ &=\ \frac{\text{(output current of time across exit plane)}}
{\text{(output particle current across exit plane)}}\\
&\ \ \ \ -\, 
\frac{\text{(input current of time across entry plane)}}
{\text{(input particle current across entry plane)}},
\end{aligned}
\end{equation}
where the exit plane is the same, or the opposite, as the 
entry plane on reflection, or on transmission, respectively.
(Similar formulas appear in \cite{R:Leavens1}, p. 110, Eqs. (26) and (27),
in \cite{R:Olkhovsky1}, Eqs.\ (28) and (29),
in \cite{R:Muga3}, Eqs. (16) and (17),
in \cite{R:Brouard1}, Eq. (61), and in \cite{R:Olkhovsky3}, Eq.\ (1).)
We therefore have that
\begin{subequations}\label{E3:eq36}
\begin{align}
\tau^{\text{trans}}_{F\leftarrow F}(z_2\leftarrow z_1)\ &=\ 
\frac{\tau^{\text{out}}_{F \leftarrow F}(z_2)}
{\mathcal{T}^\text{out}_{F \leftarrow F}(z_2)}\,-\,
\tau^{\text{in}}_F(z_1),\label{E3:eq36a}\\
\tau^{\text{refl}}_{B\leftarrow F}(z_1\leftarrow z_1)\ &=\ 
\frac{\tau^{\text{out}}_{B \leftarrow F}(z_1)}
{\mathcal{R}^\text{out}_{B \leftarrow F}(z_1)}\,-\,
\tau^{\text{in}}_F(z_1),\label{E3:eq36b}\\
\tau^{\text{trans}}_{B\leftarrow B}(z_1\leftarrow z_2)\ &=\ 
\frac{\tau^{\text{out}}_{B \leftarrow B}(z_1)}
{\mathcal{T}^\text{out}_{B \leftarrow B}(z_1)}\,+\,
\tau^{\text{in}}_B(z_2),\label{E3:eq36c}\\
\tau^{\text{refl}}_{F\leftarrow B}(z_2\leftarrow z_2)\ &=\ 
\frac{\tau^{\text{out}}_{F \leftarrow B}(z_2)}
{\mathcal{R}^\text{out}_{F \leftarrow B}(z_2)}\,+\,
\tau^{\text{in}}_B(z_2).\label{E3:eq36d}
\end{align}
\end{subequations}


\section{Discussion} \label{S:sec4}

     We have shown that the time can be construed as an observable in the
context of the Schr\"odinger equation.  We shall now undertake the discussion
of some aspects of the physical interpretation of the formalism,
and of the relation of the present work to certain other similar
investigations.

     For a state $\Psi(t,x,y,z)$ made up of a finite number
of closed-channel amplitudes that are all exponentially
decreasing with $z$ increasing, the net current of particle presence crossing
any $z$=constant surface is zero---see Eq.\ \eqref{E2:eq43}.
Zero net current means that the particle will,
in a large number of experiments,
cross negatively as often as it crosses positively.
The exponential decrease of the 
wave function is consistent with a picture of a ``virtual'' 
particle being created in the interaction region, propagating 
a small distance positively, and then falling back into the region where it
was created.  For $F$-type closed-channel states we say that
the state propagates in the positive $z$ direction, but represents
an equal flux of particle motion in positive
and negative $z$-directions.  For open-channel $F$-($B$-)type states,
propagation and global particle motion are jointly positive (negative).

     It is possible, with the given physical system, to make a measurement
of local (in $(t,x,y)$)
net transit flux of particles across a proper subset
of the $z$=constant plane.  A local measurement of either the positive
or the negative direction in $z$ of a
particle's transit involves noncommuting projection
operators, one in position space and one in wavenumber space.
That is, we would then ask two incompatible yes/no questions:  (1) Did the
particle cross the plane in a given proper $(t,x,y)$ subregion?
(2) Did the particle cross positively/negatively in
the $z$-direction?  The outcome depends on the details of an
often-repeated measurement on the system with the same input in each trial.
To be sure, taking a sufficiently large subset of the
$z$=constant plane for asking question (1) will,
to a good approximation, be close to taking all of
it, so that question (2) can be answered with negligible
inconsistency.  
Also, the question ``What is the difference
between the numbers of particles crossing positively and
particles crossing negatively 
across a small subregion of the $z$-plane?''
involves no inconsistency, and yields a
result predictable from the local wave function alone;
it is the separate
local densities of positive and of negative crossings
that depend on the measurement scheme.

     A substantial effort has been dedicated to the establishment of
a time-energy uncertainty principle---see the discussion and references
in \cite{R:Busch1}.  An uncertainty principle appears to be associated
with a positive definite metric, a requirement that we have
dropped.  It is not obviously impossible to formulate some kind of
analogous principle involving the time and its conjugate
momentum $p_t$ within the present formalism
in special circumstances, but the author has not been
successful in finding such circumstances.

     The Schr\"odinger equation can create 
spatially and temporally localized eddies of probability 
current:  even though a wave is made up entirely of a packet
of $F$-type open-channel states,
this current density can be negative in a neighborhood (see 
the examples by Kijowski and Mielnik cited in the following
paragraph) and 
therefore will be greater than one in a complementary set of the
given $z$=constant plane.  Hence,
even for a packet of only
free-particle $F$-type states this normalized particle count can
fall outside the interval $[0,1]$
on proper subsets of a $z$=constant plane, and is not a probability,
although a transition process,
as in the conventional interpretation
of quantum mechanics, has irreducible randomness and is unpredictable
in detail in contrast to a classical process.  
A further complication results from the circumstance that
this normalized particle count
is not, when closed channels are present, even globally
(i.e., across an entire $z$=constant plane) 
an algebraic sum of 
two separate currents due to forward- and backward-propagating
particles, as there can be nonzero global
interference between the
$F$- and $B$-type closed-channel contributions 
to the total current.  

     Kijowski \cite{R:Kijowski1}, \cite{R:Kijowski2}
undertook to establish a time-energy uncertainty principle by analyzing
the evolution of a Schr\"odinger wave function in a space-like direction,
and in this respect there is overlap between Kijowski's work and the present
undertaking.   Kijowski's first ``unsuccessful attempt''
(\cite{R:Kijowski1}, \S3) begins in a similar manner to
that proposed above, but his inner product law does not involve an
integral over time;  since,
as noted following Eq.\ \eqref{E2:eq37}, 
the wave functions of Eq.\ \eqref{E2:eq36}
do not satisfy the Cauchy inequality, the interference terms
in a local inner product can make the current density negative
for the superposition of two forward-propagating open channel 
states, as shown in an
example in \cite{R:Kijowski1}, \S3.  
Mielnik (\cite{R:Mielnik1}, \S5, Lemma) noted that a Schr\"odinger 
wave packet that at $t=0$ has its source entirely to the left
of $z=0$, say, could eventually give rise to probability
currents normal to the $z=0$ plane that need not be everywhere
positive.  Similarly, the integrand for
the particle current for the norm of a superposition of $F$-type 
open-channel states 
in Eq.\ \eqref{E2:eq15} need not be everywhere nonnegative.
These local negative currents all result from interference terms 
that yield zero net contribution
in the present formalism due to the integral over $t$, $x$ and $y$
in the inner product.  

     Kijowski's formalism 
is substantially different from the present one---the norm of an
$F$-type state is given in \cite{R:Kijowski1}, Eq.\ (9)---but 
in which the average time of crossing a spatial wall
for $F$-type states
nevertheless reduces to the same form (\cite{R:Kijowski1}, \S10) 
in terms of the
probability current as Eq.\ \eqref{E2:eq24}. There are discussions of
Kijowski's work in \cite{R:Muga1}, \S1.5.1, and \cite{R:Egusquiza1},
\S10.2.

    Mielnik \cite{R:Mielnik1} critiques both Kijowski's
\cite{R:Kijowski1} and Piron's \cite{R:Piron1} attempts to
establish formalisms for spacewise propagation of a wave function, 
concludes that they do not offer a solution to the
problem of defining time as an observable,
and makes no additional proposals along these lines.
Although the initial ideas of the two latter papers
resemble that of the present paper, the respective
implementations differ considerably, so we shall not attempt further
review of them here.

     Another question concerns the generalization of the 
effect of a measurement on a wave function that propagates
in both directions away from the surface on which the measurement
is performed.
Suppose in fact that, in a problem of type II, two adjacent boxes
occupy the space-like intervals $[z_1,z_2]$ and $[z_2,z_3]$, and that
a measurement is made (over $t,x,y$) at $z=z_2$,
which measurement partly ``collapses''
the wave function there.  The input at $z_2$ to both boxes can change
as a result
of the acquired information, leading in turn to a change in the overall
output at $z_1$ and $z_3$, and, due to reflections, a change in the
wave function at $z_2$ at which the measurement is made.  There is therefore
a problem of consistency, in that the measurement at $z_2$
changes the outgoing wave function  
on both sides of $z_2$, and therefore, after a reflection,
changes the ingoing wave function on both sides of $z_2$, that is, changes
the frequency of results of the measurement at $z_2$, and so on.
This problem seems analogous to the ``grandfather paradox''
(see \cite{R:Nahin1}, Ch.\ 4)
of the influence of a physical system
with itself between two different $t$=constant surfaces,
when reflection as well as transmission of signals along the
evolution coordinate occurs.  
No successful attempt at analysis of this class of measurement
problems is known to the author.

     We re{\"e}mphasize  
that it is the Schr\"odinger equation that is taken as fundamental
in the present argument, and that a theory of measurement,
a probability interpretation, 
and an uncertainty principle, are all
presumed to be derivative ideas that may require
alterations from their conventional forms
in order to bring them into concord with the body of formalism presented here.
We have assumed that not just $\psi^\ast\psi$, but also the 
spatial components of the probability four-current of Sec.\ 2,
are measurable quantities.  Moreover, we assume that the flux density
of a physical quantity represented by an operator $\omega$
is given by the four-current of Eqs.\ \eqref{E2:eq20}.
In classical terms, the total fluxes amount to normalized counts
of particles crossing a given oriented three-dimensional
surface in space-time, weighted by the time of crossing or some other
such quantity, and also weighted positively or negatively
according as the particle crosses positively or negatively
when it transits the given surface.  Note that 
in experimental trials the particle is
presumed always
to transit positively across a segment of a $t$=constant 
surface---Galilean geometry singles out $t$=constant planes from all 
other planes in space-time and permits special treatment for these cases.  

     To recapitulate in other words, 
we assert that the evolution of a Schr{\"o}dinger wave function
in a spatial direction does not generally admit of description
in terms of probability amplitudes.
The claim is rather that the wave function in type II problems
permits only the computation of certain 
expectation values, that is, average results of many
repeated experiments with the same input signal, but such that
there exists no underlying distribution of nonnegative quantities,
analogous to $\psi^\ast\psi$,  that 
accounts for the results.  We advocate the non-introduction of 
the term ``negative probabilities'' as there
seems to result a decrease in physical clarity thereby; 
instead, a kind of random behavior more general 
than that which can be characterized by probabilities is
entailed. Khrennikov, in \cite{R:Khrennikov1}, Ch.\ III.2
and references given therein,
describes what he calls ``signed `probabilistic' measures'',
which could serve as a classical analog of the local norm
of Eq.\ \eqref{E2:eq13};  see also \cite{R:Kolmogorov1}, \S 34.
We infer that the conventional, probability 
interpretation of nonrelativistic quantum mechanics
should be subordinated to an interpretation involving observable 
stochastic currents of particle presence, and, more comprehensively,
observable 
stochastic currents of other physical quantities as temporal position, 
spatial position, energy, momentum, and so on.  
This ``particle current'' interpretation
of the formalism can describe systems of both
types I and II;  the 
usual probability interpretation then applies 
in problems of type I and other special cases.

     Although what appears to be a mathematically
consistent formalism has been constructed herein, 
and a preliminary physical interpretation advanced,
many questions along these lines need to be addressed,
and consistency with experimental tests established,
before the proposal can with confidence be regarded as a
physical theory.

     The above limitations notwithstanding, the formalism 
proposed herein has obtained results that agree to an extent
with some special results previously derived, and has
secured results that would be difficult
to obtain by other published methods of analysis:
for example, a generic expression Eqs.\ \eqref{E3:eq23}--\eqref{E3:eq28}
for the average dwell time for a particle 
reflecting from or passing through a time-dependent barrier.



\begin{thebibliography}{10}

\bibitem{R:Neumann2}
J.~von Neumann.
\newblock {\em Mathematische Grundlagen der Quantenmechanik}.
\newblock Springer, Berlin, Germany, 1932.

\bibitem{R:Neumann1}
J.~von Neumann.
\newblock {\em Mathematical Foundations of Quantum Mechanics}.
\newblock Princeton U. Pr., Princeton, NJ, USA, 1955.
\newblock (English translation of Ref.\ \cite{R:Neumann2}).

\bibitem{R:Pauli2}
W.~Pauli.
\newblock Allgemeine prinzipien der wellenmechanik.
\newblock volume XXIV, Part 1 of {\em Handbuch der Physik}. Springer, Berlin,
  Germany, 1933.

\bibitem{R:Pauli1}
W.~Pauli.
\newblock {\em General Principles of Quantum Mechanics}.
\newblock Springer-Verlag, Berlin, Germany, 1980.
\newblock (English translation of Ref.\ \cite{R:Pauli2}).

\bibitem{R:Peres1}
A.~Peres.
\newblock {\em Quantum Theory: Concepts and Methods}.
\newblock Kluwer Academic, Dordrecht, The Netherlands, 1995.

\bibitem{R:Omnes1}
R.~Omn{\`e}s.
\newblock {\em The Interpretation of Quantum Mechanics}.
\newblock Princeton U. Press, Princeton, NJ, USA, 1994.

\bibitem{R:Sakurai1}
J.~J. Sakurai.
\newblock {\em Modern Quantum Mechanics}.
\newblock Addison-Wesley, Reading, MA, USA, revised edition, 1994.

\bibitem{R:Capasso1}
F.~Capasso, K.~Mohammed, and A.~Y. Cho.
\newblock {\em IEEE J. Quant. Electr.}, QE-22:1853, 1986.

\bibitem{R:Ferry1}
D.~K. Ferry.
\newblock {\em Transport in nanostructures}.
\newblock Cambridge U. Pr., Cambridge, UK, 1997.

\bibitem{R:Jonson1}
M.~Jonson.
\newblock In D.~K. Ferry and C.~Jacoboni, editors, {\em Quantum Transport in
  Semiconductors}, chapter~10. Plenum, New York, NY, 1992.

\bibitem{R:Jauho1}
A.~P. Jauho.
\newblock In D.~K. Ferry and C.~Jacoboni, editors, {\em Quantum Transport in
  Semiconductors}, chapter~9. Plenum, New York, NY, 1992.

\bibitem{R:Jauho2}
A.~P. Jauho.
\newblock In J.~Shah, editor, {\em Hot Carriers in Semiconductor
  Nanostructures}, chapter II.4. Academic, Boston, MA, 1992.

\bibitem{R:Mizuta1}
H.~Mizuta and T.~Tanoue.
\newblock {\em The Physics and applications of resonant tunnelling diodes}.
\newblock Cambridge U. Press, Cambridge, UK, 1995.

\bibitem{R:Hauge1}
E.~H. Hauge and J.~A. St{\o}vening.
\newblock {\em Rev. Mod. Phys.}, 61:917, 1989.

\bibitem{R:Olkhovsky1}
V.~S. Olkhovsky and E.~Recami.
\newblock {\em Phys. Reports}, 214:339, 1991.

\bibitem{R:Landauer1}
R.~Landauer and Th. Martin.
\newblock {\em Rev. Mod. Phys.}, 66:217, 1994.

\bibitem{R:Chiao1}
R.~Y. Chiao and A.~M. Steinberg.
\newblock {\em Progress in Optics}, XXXVII:345, 1997.

\bibitem{R:Mugnai1}
D.~Mugnai, A.~Ranfagni, and L.~S. Schulman, editors.
\newblock {\em Tunneling and its implications}.
\newblock World Scientific, River Edge, NJ, 1997.

\bibitem{R:Damborenea1}
J.~A. Damborenea, I.~L. Egusquiza, G.~C. Hegerfeldt, and J.~G. Muga.
\newblock {\em Phys. Rev. A}, 66:052104, 2002.

\bibitem{R:Muga0}
J.~G. Muga, R.~S. Mayato, and I.~L. Egusquiza, editors.
\newblock {\em Time in Quantum Mechanics}.
\newblock Lecture Notes in Physics M72. Springer, Berlin, Germany, 2002.

\bibitem{R:Lanczos1}
C.~Lanczos.
\newblock {\em The Variational Principles of Mechanics}.
\newblock Dover, New York, NY, USA, 4th edition, 1986.

\bibitem{R:Friedman1}
A.~Friedman.
\newblock {\em Partial differential equations of parabolic type}.
\newblock Prentice-Hall, Englewood Cliffs, NJ, 1964.

\bibitem{R:Gohberg1}
I.~Gohberg, P.~Lancaster, and L.~Rodman.
\newblock {\em Matrices and Indefinite Scalar Products}.
\newblock Birkh{\"{a}}user, Basel, Switzerland, 1983.

\bibitem{R:Kijowski1}
J.~Kijowski.
\newblock {\em Rep. Math. Phys.}, 6:361, 1974.

\bibitem{R:Piron1}
C.~Piron.
\newblock {\em C. R. Acad. Sc. Paris}, A286:713, 1978.

\bibitem{R:Mielnik1}
B.~Mielnik.
\newblock {\em Found. Phys.}, 24:1113, 1994.

\bibitem{R:Hahne1}
G.~E. Hahne.
\newblock {\em J. Phys. A}, 35:7101, 2002.

\bibitem{R:Schiff1}
L.~I. Schiff.
\newblock {\em Quantum Mechanics}.
\newblock McGraw-Hill, New York, NY, USA, 3rd edition, 1968.

\bibitem{R:Goldstein1}
H.~Goldstein.
\newblock {\em Classical Mechanics}.
\newblock Addison Wesley, Reading, MA, USA, 2nd edition, 1980.

\bibitem{R:Miller1}
W.~H. Miller.
\newblock In D.~C. Clary, editor, {\em The Theory of Chemical Reaction
  Dynamics}, page~27. Dordrecht, The Netherlands, 1986.

\bibitem{R:Muga2}
J.~G. Muga.
\newblock Characteristic times in one-dimensional scattering.
\newblock In J.~G. Muga, R.~S. Mayato, and I.~L. Egusquiza, editors, {\em Time
  in Quantum Mechanics}, Lecture Notes in Physics M72, chapter~2. Springer,
  Berlin, Germany, 2002.

\bibitem{R:Nussenzweig1}
H.~M. Nussenzweig.
\newblock {\em Phys. Rev. A}, 62:042107, 2000.

\bibitem{R:Jaworski1}
W.~Jaworski and D.~M. Wardlaw.
\newblock {\em Phys. Rev. A}, 37:2843, 1988.

\bibitem{R:Jaworski2}
W.~Jaworski and D.~M. Wardlaw.
\newblock {\em Phys. Rev. A}, 38:5404, 1988.

\bibitem{R:Jaworski3}
W.~Jaworski and D.~M. Wardlaw.
\newblock {\em Phys. Rev. A}, 40:6210, 1989.

\bibitem{R:Jaworski4}
W.~Jaworski and D.~M. Wardlaw.
\newblock {\em Phys. Rev. A}, 45:292, 1992.

\bibitem{R:Heisenberg1}
W.~Heisenberg.
\newblock {\em Nucl. Phys.}, 4:532, 1957.

\bibitem{R:Nagy1}
K.~L. Nagy.
\newblock {\em State Vector Spaces with Indefinite Metric in Quantum Field
  Theory}.
\newblock P. Noordhoff, Groningen, The Netherlands, 1966.

\bibitem{R:MacLane1}
S.~MacLane and G.~Birkhoff.
\newblock {\em Algebra}.
\newblock Chelsea, New York, NY, USA, 3rd edition, 1988.

\bibitem{R:Goldberger1}
M.~L. Goldberger and K.~M. Watson.
\newblock {\em Collision Theory}.
\newblock J. Wiley, New York, NY, USA, 1964.

\bibitem{R:Smith1}
F.~T. Smith.
\newblock {\em Phys. Rev.}, 118:349--356, 1960.

\bibitem{R:Smith2}
F.~T. Smith.
\newblock {\em Phys. Rev.}, 119:2098, 1960.

\bibitem{R:Leavens1}
C.~R. Leavens.
\newblock Tunneling and its implications.
\newblock World Scientific, River Edge, NJ, 1997.

\bibitem{R:Muga3}
J.~G. Muga, S.~Brouard, and R.~Sala.
\newblock {\em Phys. Lett. A}, 167:24, 1992.

\bibitem{R:Brouard1}
S.~Brouard, R.~Sala, and J.~G. Muga.
\newblock {\em Phys. Rev. A}, 49:4312, 1994.

\bibitem{R:Olkhovsky3}
V.~S. Olkhovsky, E.~Recami, and J.~Jakiel.
\newblock {\em online arXiv:quant-ph/0102007}, 2001.

\bibitem{R:Busch1}
P.~Busch.
\newblock The time-energy uncertainty relation.
\newblock In J.~G. Muga, R.~S. Mayato, and I.~L. Egusquiza, editors, {\em Time
  in Quantum Mechanics}, Lecture Notes in Physics M72, chapter~3. Springer,
  Berlin, Germany, 2002.

\bibitem{R:Kijowski2}
J.~Kijowski.
\newblock {\em Phys. Rev. A}, 59:897, 1999.

\bibitem{R:Muga1}
J.~G. Muga, R.~S. Mayato, and I.~L. Egusquiza.
\newblock In J.~G. Muga, R.~S. Mayato, and I.~L. Egusquiza, editors, {\em Time
  in Quantum Mechanics}, Lecture Notes in Physics M72, chapter~1. Springer,
  Berlin, Germany, 2002.

\bibitem{R:Egusquiza1}
I.~L. Egusquiza, J.~G. Muga, and A.~D. Baute.
\newblock ``{S}tandard'' quantum mechanical approach to times of arrival.
\newblock In J.~G. Muga, R.~S. Mayato, and I.~L. Egusquiza, editors, {\em Time
  in Quantum Mechanics}, Lecture Notes in Physics M72, chapter~10. Springer,
  Berlin, Germany, 2002.

\bibitem{R:Nahin1}
P.~J. Nahin.
\newblock {\em Time machines}.
\newblock Springer, New York, NY, 2nd edition, 1999.

\bibitem{R:Khrennikov1}
A.~Khrennikov.
\newblock {\em Interpretations of Probability}.
\newblock VSP, Utrecht, The Netherlands, 1999.

\bibitem{R:Kolmogorov1}
A.~N. Kolmogorov and S.~V. Fomin.
\newblock {\em Introductory Real Analysis}.
\newblock Prentice-Hall, Englewood Cliffs, NJ, USA, 1970.

\end{thebibliography}
\end{document}